\documentclass[12pt]{article}
\newlength{\abstractwidth}
\usepackage{amsmath}
\usepackage{amsfonts}
\usepackage{amssymb}
\usepackage{latexsym}

\usepackage{epsf}
\usepackage{color}
\usepackage{graphicx}
\usepackage{tikz}
\usepackage{dsfont}
\usepackage{subfigure}

\usepackage{hyperref}
\definecolor{darkred}{rgb}{0.8,0.1,0.1}
\hypersetup{colorlinks=true, linkcolor=darkred, citecolor=blue, linktoc=page}

\usetikzlibrary{arrows,shapes,positioning}
\usetikzlibrary{decorations.markings}
\usepackage[rightcaption]{sidecap}
\tikzstyle arrowstyle=[scale=1]
\tikzstyle directed=[postaction={decorate,decoration={markings,
    mark=at position .65 with {\arrow[arrowstyle]{stealth}}}}]
\tikzstyle reverse directed=[postaction={decorate,decoration={markings,
    mark=at position .65 with {\arrowreversed[arrowstyle]{stealth};}}}]
\usetikzlibrary{positioning}

\renewcommand{\thefootnote}{\fnsymbol{footnote}}
\renewcommand{\thanks}[1]{\footnote{#1}}
\newcommand{\starttext}{
\setcounter{footnote}{0}
\renewcommand{\thefootnote}{\arabic{footnote}}}

\newcommand{\bea}{\begin{eqnarray}}
\newcommand{\eea}{\end{eqnarray}}
\newcommand{\be}{\begin{eqnarray}}
\newcommand{\ee}{\end{eqnarray}}
\newcommand{\<}{\langle}
\renewcommand{\>}{\rangle}

\newcommand{\tx}{\text}
\def\cC{{\cal C}}
\def\cD{{\cal D}}

\def\cK{{\cal K}}
\def\cL{{\cal L}}

\def\cN{{\cal N}}
\def\cO{{\cal O}}

\def\cW{{\cal W}}

\def\ZZ{{\mathbb Z}}
\def\RR{{\mathbb R}}

\def\Tr{{\rm Tr}}
\def\det{{\rm det \,}}

\def\half{{1\over 2}}
\def\p{\partial}

\def\a{\alpha}
\def\b{\beta}
\def\g{\gamma}

\def\eps{\epsilon}
\def\f{\varphi}

\def\ep{\varepsilon}

\def\s{\sigma}
\def\m{\mu}
\def\){\right)}
\def\({\left( }
\def\]{\right] }
\def\[{\left[ }
\def\no{\nonumber}

\makeatletter\def\l@subsubsection#1#2{}%
\makeatother

\begin{document}
\starttext
\setcounter{footnote}{0}

\vskip 0.3in

\begin{center}

{\Large \bf Mass deformations of 5d SCFTs via holography}

\vskip 0.4in

{\large   Michael Gutperle$^a$, Justin Kaidi$^a$  and Himanshu Raj$^{a,b}$} 

\vskip .2in

$^a$ {\it Mani L. Bhaumik Institute for Theoretical Physics}\\
{ \it Department of Physics and Astronomy }\\
{\it University of California, Los Angeles, CA 90095, USA}\\[0.5cm]

$^b${\it  SISSA and INFN} \\ 
 {\it Via Bonomea 265; I 34136 Trieste, Italy} \\
 
 \vskip .2in
 
\begin{abstract}
 
Using six-dimensional Euclidean $F(4)$ gauged supergravity we construct  a holographic renormalization group flow for  a CFT on  $S^5$. Numerical solutions to the BPS equations are obtained and the free energy of the theory on $S^5$ is determined holographically by calculation of the renormalized on-shell supergravity action. In the process, we deal with subtle issues such as holographic renormalization and addition of finite counterterms. We then propose a candidate field theory dual to these solutions. This tentative dual is a supersymmetry-preserving  deformation of the strongly-coupled non-Lagrangian SCFT derived from the D4-D8 system in string theory.  In the IR, this theory is a mass deformation of a $USp(2N)$ gauge theory. A localization calculation of the free energy is performed for this IR theory, which for reasonably small values of the deformation parameter is found to have the same qualitative behaviour as the holographic free energy.

\end{abstract}
\end{center}

\baselineskip=16pt
\setcounter{equation}{0}
\setcounter{footnote}{0}

\newpage
\tableofcontents
\newpage
\section{Introduction}
\setcounter{equation}{0}
\label{sec1}

Despite being non-renomalizable, five-dimensional supersymmetric field theories have a history of study via string and M-theory
\cite{Seiberg:1996bd,Morrison:1996xf,Intriligator:1997pq}. A plethora of five-dimensional gauge theories can be  realized by utilizing  brane constructions in type IIA  \cite{Brandhuber:1999np,Bergman:2012kr} as well as type IIB string theory \cite{Aharony:1997ju,Aharony:1997bh,DeWolfe:1999hj}. In contrast to theories in other dimensions, five-dimensional superconformal field theories (SCFTs) have a unique superalgebra  $F(4)$ \cite{Nahm:1977tg,Kac:1977em,Shnider:1988wh} with $SO(2,5)$ conformal symmetry, $SU(2)_R$ R-symmetry, and sixteen supercharges (eight Poincare and eight conformal supercharges).

Though difficult to study directly, holography can be used to study five-dimensional SCFTs in the large $N$ limit.
Supergravity solutions containing an AdS$_6$ factor had previously been found in massive type IIA supergravity \cite{Brandhuber:1999np,Bergman:2012kr,Passias:2012vp} as well as in type IIB supergravity \cite{Lozano:2012au,Lozano:2013oma,Kelekci:2014ima}. 
In the last year,\footnote{For earlier work in this direction, see \cite{Apruzzi:2014qva,Kim:2015hya,Kim:2016rhs}.} new type IIB supergravity solutions were found using an ansatz with AdS$_6\times S^2$ warped over a two-dimensional Riemann surface $\Sigma$ with boundary \cite{DHoker:2016ujz,DHoker:2016ysh,DHoker:2017mds,Gutperle:2017tjo}. Aside from isolated points on the boundary of $\Sigma$, these solutions are completely regular. At these isolated points the harmonic functions which determine the solutions have poles. These poles can be given a physical interpretation as the remnants of semi-infinite $(p,q)$ five-branes resulting from the conformal limit of $(p,q)$ five-brane webs  \cite{Aharony:1997ju,Aharony:1997bh,DeWolfe:1999hj}. However, these solutions are technically involved.

A simpler setting for AdS$_6$/CFT$_5$ duality is given by six-dimensional $F(4)$ gauged supergravity. $F(4)$ gauged supergravity was first constructed in \cite{Romans:1985tw}. The theory can be coupled to any number of six-dimensional vector multiplets, with the resulting Lagrangian, supersymmetry transformations, and possible gaugings found in \cite{Andrianopoli:2001rs}. These theories admit supersymmetric AdS$_6$ vacua, and determining the spectrum of linearized supergravity fluctuations dual to primary operators is straightforward \cite{Ferrara:1998gv,DAuria:2000afl,Karndumri:2016ruc}. For some additional work on the use of $F(4)$ gauged supergravity in holography, see e.g. \cite{Karndumri:2012vh,Karndumri:2014lba,Alday:2014rxa,Alday:2014fsa,Hama:2014iea}.
To our knowledge, it is not yet known how to lift general solutions of six-dimensional gauged supergravity to ten dimensions, and hence a microscopic understanding of the CFT described by the AdS vacua is still lacking.  On the other hand, obtaining and studying solutions for the six-dimensional theory is relatively simple, and general lessons also applicable to more complicated theories can be learned. An example of such solutions is the 6d supersymmetric Janus solution constructed recently in \cite{Gutperle:2017nwo}.

In this paper, we will be interested in studying certain deformations of 5d SCFTs via holography. The superconformal symmetry of a SCFT can be broken by turning on relevant operators, some of which may keep (some) Poinc\'are supersymmetries unbroken. Well known cases of such deformations include the $\cN=2^*$  and $\cN=1^*$ theories obtained by mass deformations of $\cN=4$ super Yang-Mills. A systematic classification of operators which break superconformal symmetry but leave Poincare supersymmetry unbroken was recently obtained in \cite{Cordova:2016xhm}. In order to make use of localization results, we will furthermore be interested in deformed SCFTs on $S^5$.  Conformal field theories defined on $\RR^d$ can be put on other conformally flat manifolds such as the $d$-dimensional sphere in a unique fashion. However, for non-conformal theories this is not the case, though for many theories it is possible turn on additional terms in the Lagrangian which preserve supersymmetry on the curved space. For $\cN=2^*$ these terms were found in \cite{Pestun:2007rz} and for gauge theories on $S^5$ such terms were given in \cite{Hosomichi:2012ek,Kallen:2012va}.

In the context of the AdS/CFT correspondence, such deformations on spheres have been studied for four-dimensional $\cN=2^*$  \cite{Bobev:2013cja}, four-dimensional $\cN=1^*$ \cite{Bobev:2016nua}, and three-dimensional ABJM theories \cite{Freedman:2013ryh}.
The method used to study these theories holographically is as follows. For a field theory in $d$-dimensions, one considers a gauged supergravity with an AdS$_{d+1}$ vacuum corresponding to the undeformed superconformal field theory. The ansatz for the metric corresponding to the deformed theory is given by a Euclidean RG-flow/domain wall, where a $d$-dimensional sphere is warped over a one-dimensional holographic direction. The scalars which are dual to the mass deformations, as well as the additional terms which are necessary for preserving supersymmetry on the sphere, are sourced in the UV. 
The preservation of supersymmetry in the supergravity demands the vanishing of fermionic supersymmetry variations and provides first-order flow equations for the scalars. The integrability conditions for the gravitino variation determine the metric. For generic scalar sources, the flow will lead to a singular solution, but demanding that the sphere closes off smoothly in the IR provides relations among  the UV sources and leads to a nonsingular supersymmetric RG flow. 
Using holographic renormalization, the free energy of the theory on the sphere is determined by calculating the renormalized on-shell action of the supergravity solutions. The continuation of the supergravity theory from Lorentzian to Euclidean signature, the precise mapping of supergravity fields to field theory operators, and the choice of finite counterterms preserving supersymmetry are among the subtle issues which the papers  \cite{Bobev:2013cja,Bobev:2016nua,Freedman:2013ryh} address in five- and four-dimensional gauged supergravity.

The goal of this paper is to apply these techniques to matter-coupled six-dimensional gauged supergravity \cite{Romans:1985tw,Andrianopoli:2001rs} in order to study mass deformations of a five-dimensional SCFT on $S^5$. 
The structure of  the paper is as follows: In section \ref{sec2},  we review features of the Lorentzian matter-coupled $F(4)$ gauged supergravity theory. In section
\ref{sec3}, we discuss the continuation of the supergravity to Euclidean signature and construct the ansatz describing the RG flow on $S^5$. Vanishing of the fermionic variations leads to the Euclidean BPS equations. We solve these equations numerically and obtain a one parameter family of smooth solutions. In section \ref{sec4}, we use holographic renormalization to evaluate the on-shell action as a function of the mass parameter. In the process, we deal with the subtle issue of identification of finite counterterms needed to preserve supersymmetry on $S^5$. In section \ref{fieldtheory}, we compare the holographic sphere free energy with the corresponding result obtained via localization in the large $N$ limit of a $USp(2N)$ gauge theory with one massless hypermultiplet in the antisymmetric representation and one massive hypermultiplet in the fundamental representation of the gauge group. In section \ref{sec5}, we close with a discussion of our results and future directions for research.

%%%%%%%%%%%%%%%%%%%%%%%%%%%%%%%%%%%%%%%%%%%%
%%%%%%%%%%%%%%%%%%%%%%%%%%%%%%%%%%%%%%%%%%%%
\section{Lorentzian matter-coupled  $F(4)$ gauged supergravity }
\setcounter{equation}{0}
\label{sec2}
%%%%%%%%%%%%%%%%%%%%%%%%%%%%%%%%%%%%%%%%%%%%
%%%%%%%%%%%%%%%%%%%%%%%%%%%%%%%%%%%%%%%%%%%%

The theory of matter-coupled $F(4)$ gauged supergravity was first studied in \cite{Andrianopoli:2001rs,DAuria:2000afl}, with some applications and extensions given in \cite{Karndumri:2016ruc,Karndumri:2012vh,Karndumri:2014lba}. Below we present a short review of this theory, similar to that given in \cite{Gutperle:2017nwo}.

\subsection{The bosonic Lagrangian}
\label{sec2.1}
We begin by recalling the field content of the 6-dimensional supergravity multiplet, 
\bea
(e^a_\mu,\, \psi^A_\mu,\, A^\a_\mu, \,B_{\mu \nu}, \,\chi^A, \,\sigma)
\eea
The field $e^a_\mu$ is the 6-dimensional frame field, with spacetime indices denoted by $\{\mu, \nu\}$ and local Lorentz indices denoted by $\{a,b\}$.
The field $\psi^A_\mu$ is the gravitino with the index $A ,B= 1,2$ denoting the fundamental representation of the gauged $SU(2)_R$ group. The supergravity multiplet contains four vectors $A_\mu^\a$ labelled by the index $\a = 0, \dots 3$. It will often prove useful to split $\a = (0,r)$ with $r = 1, \dots, 3$ an $SU(2)_R$ adjoint index. Finally, the remaining fields consist of a two-form $B_{\mu \nu}$, a spin-${1 \over 2}$ field $\chi^A$, and the dilaton $\sigma$.

The only allowable matter in the $d=6$, $\cN=2$ theory is the vector multiplet, which has the following field content 
\bea
(A_\mu, \,\lambda_A, \,\phi^\a)^I
\eea
where $I = 1, \dots, n$ labels the distinct matter multiplets included in the theory. The presence of the $n$ new vector fields $A_\mu^I$ allows for the existence of a further gauge group $G_+$ of dimension $\mathrm{dim} \,G_+ = n$, in addition to the gauged $SU(2)_R$ R-symmetry. The presence of this new gauge group contributes an additional parameter to the theory, in the form of a coupling constant $\lambda$. Throughout this section, we will denote the structure constants of the additional gauge group $G_+$ by $C_{I J K}$.  However, these will play no role in what follows, since we will be restricting to the case of only a single vector multiplet $n=1$, in which case $G_+=U(1)$.

In (half-)maximal supergravity, the dynamics of the $4n$ vector multiplet scalars $\phi^{\a I}$ is given by a non-linear sigma model with target space $G/K$; see e.g. \cite{Samtleben:2008pe}. The group $G$ is the global symmetry group of the theory, while $K$ is the maximal compact subgroup of $G$. As such, in the Lorentzian case the target space is identified with the following coset space, 
\bea\label{coseta}
{\cal M}= {SO(4,n) \over SO(4) \times SO(n) }\times SO(1,1)
\eea
where the second factor corresponds to the scalar $\sigma$ which is already present in the gauged supergravity without added matter. In the particular case of $n=1$, explored here and in \cite{Gutperle:2017nwo}, the first factor is nothing but four-dimensional hyperbolic space $\mathbb{H}_4$. When we analytically continue to the Euclidean case, it will prove very important that we analytically continue the coset space as well, resulting in a dS$_4$ coset space. This will be discussed more in the following section.

In both the Lorentzian and Euclidean cases, a convenient way of formulating the coset space non-linear sigma model is to have the scalars $\phi^{\a I}$ parameterize an element $L$ of $G$. The so-called coset representative $L$ is an $(n+4)\times(n+4)$ matrix with matrix elements $L^\Lambda_{\,\,\,\,\Sigma}$, for  $\Lambda, \Sigma = 1, \dots n+4$. Using this representative, one may construct a left-invariant 1-form, 
\bea
L^{-1} d L \in \mathfrak{g}
\eea
where $\mathfrak{g} = \mathrm{Lie}(G)$. To build a $K$-invariant kinetic term from the above, we decompose
\bea
L^{-1} d L = Q + P
\eea
where $Q \in \mathfrak{k} = \mathrm{Lie}(K)$ and $P$ lies in the complement of $\mathfrak{k}$ in $\mathfrak{g}$.  Explicitly, the coset vielbein forms are  given by, 
\bea
\label{cosetvielbeins}
P^I_{\,\,\,\,\a} = \left(L^{-1}\right)^I_{\,\,\,\,\Lambda} \left(d L^\Lambda_{\,\,\,\,\a} + f^\Lambda_{\,\,\,\,\Gamma \Pi} A^\Gamma L^\Pi_{\,\,\,\,\a} \right)
\eea
where the $f_{\Lambda \Sigma}^{\,\,\,\,\,\,\,\,\Gamma}$ are structure constants of the gauge algebra, i.e.
\bea
[T_\Lambda, T_\Sigma] =  f_{\Lambda \Sigma}^{\,\,\,\,\,\,\,\,\Gamma} \, T_\Gamma
\eea
We may then use $P$ to build the kinetic term for the vector multiplet scalars as, 
\bea
\cL_{\text{coset}} = -{1\over 4} e P_{I \a \mu} P^{I \a \mu}
\eea
where $e = \sqrt{|\mathrm{det} \,g|}$ and we've defined $P_\mu^{I \a} = P_i^{I \a} \p_\mu \phi^i,$ for $i = 0, \dots, 4n-1$.  With this formulation for the coset space non-linear sigma model, we may now write down the full bosonic Lagrangian of the theory. We will be interested in the case in which only the metric and the scalars are non-vanishing. In this case the Lorentzian theory is given by
 \bea
 \label{Lagrangian}
 e^{-1} \cL = -{1 \over 4} R + \p_\mu \sigma \p^\mu \sigma - {1\over 4} P_{I \a \mu} P^{I \a \mu} - V
 \eea
 with the scalar potential $V$ given by 
 \bea
 \label{scalarpot}
V &=& -e^{2 \sigma} \left[ {1 \over 36} A^2 + {1 \over 4}B^i B_i + {1 \over 4} (C^I_t C_{I t} + 4 D^I_t D_{I t}) \right] + m^2 e^{-6 \sigma} \cN_{00} 
\no\\
&\vphantom{.}& \hspace{0.5in} - m e^{- 2 \sigma} \left[{2 \over 3} A L_{00} - 2 B^i L_{0 i} \right]
\eea 
The scalar potential features the following quantities, 
\bea
A &=& \eps^{r s t} K_{r s t} \hspace{1 in}\,\,B^r\, = \,\eps^{r s t} K_{s t 0}
\no\\
C_I^t &=& \eps^{t r s} K_{r I s} \hspace{1 in} D_{I t}\,\, = \,\, K_{0 I t}
\eea
with the so-called ``boosted structure constants" $K$ given by,
\bea
K_{r s \a} &=& g \,\eps_{\ell m n} L^{\ell}_{\,\,\,\, r} (L^{-1})_s^{\,\,\,\, m} L^n_{\,\,\,\,\a} + \lambda \,C_{I J K} L^I_{\,\,\,\,r} (L^{-1})_s^{\,\,\,\,J}L^K_{\,\,\,\,\,\a}
\no\\
K_{\a I t} &=& g\, \eps_{\ell m n} L^{\ell}_{\,\,\,\, \a} (L^{-1})_I^{\,\,\,\, m} L^n_{\,\,\,\,t} + \lambda \,C_{M J K} L^M_{\,\,\,\,\,\a} (L^{-1})_I^{\,\,\,\,J}L^K_{\,\,\,\,\,t}
\eea
We remind the reader that $r,s,t = 1, 2, 3$ are obtained from splitting the index $\a$ into a 0 index and an $SU(2)_R$ adjoint index.  Also appearing in the Lagrangian is $\cN_{00}$, which is the $00$ component of the matrix 
\bea
\cN_{\Lambda \Sigma} = L_\Lambda^{\,\,\,\,\a}\left( L^{-1}\right)_{\a \Sigma} -  L_\Lambda^{\,\,\,\,I}\left( L^{-1}\right)_{I \Sigma} 
\eea

\subsection{Supersymmetry variations}
We now review the supersymmetry variations for the fermionic fields in the Lorentzian theory. In the following section, we will discuss the continuation of this theory to Euclidean signature, which is complicated by the necessary modification of the symplectic Majorana condition imposed on the spinor fields.

In order to write the fermionic variations, it is first necessary to introduce a matrix $\g^7$ defined as 
\bea
\g^7=i\g^0\g^1\g^2\g^3\g^4\g^5
\eea
and satisfying $(\g^7)^2=-\mathds{1}$. With this, the supersymmetry transformations of the fermions in the Lorentzian case can be given as
\bea
\label{susyvariationsmink}
\delta \chi_A &=& {i \over 2} \g^\m \p_\m \sigma \ep_A + N_{AB} \ep^B 
\no\\\no\\
\delta \psi_{A \m} &=& \cD_\m \ep_A + S_{AB} \g_\m \ep^B
\no\\\no\\
\delta \lambda^I_A &=& i \hat P^I_{r i} \sigma^r_{AB} \p_\m \phi^i \g^\m \ep^B -i \hat P^I_{0 i} \epsilon_{AB} \p_\m \phi^i \g^7 \g^\m \ep^B + M^I_{AB} \ep^B
\eea
where we have defined
\begin{eqnarray}
\label{SMNvars}
S_{AB}&=&\!\frac{i}{24}[Ae^{\sigma}\! +\!
6me^{-3\sigma}(L^{-1})_{00}]\varepsilon_{AB}\! -\!
\frac{i}{8}[B_te^{\sigma}-2me^{-3\sigma}(L^{-1})_{t0}]\gamma^7\sigma^t_{AB}\no
\\\no\\
N_{AB}&=&\!\frac{1}{24}[Ae^{\sigma}\! -\!
18me^{-3\sigma}(L^{-1})_{00}]\varepsilon_{AB}\! +\!
\frac{1}{8}[B_te^{\sigma}\! +\! 6me^{-3\sigma}(L^{-1})_{t0}]\gamma^7\sigma^t_{AB}\no\\
\no\\
M^I_{AB}&=&\!(-C^I_{~t}+2i\gamma^7D^I_{~t})e^{\sigma}\sigma^t_{AB}-
2me^{-3\sigma}(L^{-1})^I_{\ \ 0}\g^7\varepsilon_{AB}, \label{mgM}
\end{eqnarray}
In the above, the matrix $\sigma^r_{AB}$ defined as $\sigma^r_{AB}\equiv\sigma^{rC}_{~~B}\varepsilon_{CA}$ is symmetric in $A,B$. For more details, see  our previous paper \cite{Gutperle:2017nwo}.

\subsection{Mass deformations}
\label{sec2massdef}
In the following, we consider the coset (\ref{coseta}) with $n=1$, i.e. a single vector multiplet. The coset representative is expressed in terms of four scalars $\phi^i, i=0,1,2,3$ via
\bea
\label{cosetrep}
L=\prod_{i=0}^3e^{\phi^i K^i}
\eea
where $K^i$ are the non compact generators of $SO(4,1)$; see \cite{Gutperle:2017nwo} for details. Note that  $\phi^0$ is an $SU(2)_R$ singlet, while the other three scalars $\phi^r$ form an $SU(2)_R$ triplet. 
The scalar potential for this specific case can be obtained from (\ref{scalarpot}) and takes the following form
\begin{align}\label{scalpota}
V(\sigma,\phi^i) =&- g^2 e^{2 \sigma }+\frac{1}{8} m e^{-6 \sigma } \bigg[-32 g e^{4 \sigma } \cosh \phi^0 \cosh \phi^1 \cosh \phi^2 \cosh \phi^3+8 m \cosh ^2\phi^0
\no\\
&+m \sinh ^2 \phi^0 \bigg(-6+8 \cosh ^2\phi^1 \cosh ^2 \phi^2 \cosh (2 \phi^3)+\cosh (2 (\phi^1-\phi^2))\no\\
&+\cosh (2 (\phi^1+\phi^2))+2 \cosh (2 \phi^1)+2 \cosh (2 \phi^2)\bigg)\bigg]
\end{align}
The supersymmetric AdS$_6$ vacuum is given by setting $g=3m$ and setting all scalars to vanish. The masses of the  linearized  scalar fluctuation around the AdS vacuum  determine the dimensions of the dual scalar operators in the SCFT via
\be
\label{usualthing}
m^2l^2= \Delta(\Delta-5)
\ee
where $l$ is the curvature radius of the AdS$_6$ vacuum. For the scalars at hand, one finds
\bea
m_\sigma^2 l^2 = -6 \hspace{0.8 in} m_{\phi^0}^2 l^2 = -4 \hspace{0.8 in} m_{\phi^r}^2 l^2 = -6\,\,,\,\,\,\,r=1,2,3
\eea
Hence the dimensions of the dual operators are 
\bea\label{opedic}
\Delta_{{\cal O}_\sigma} = 3, \hspace{0.8 in} \Delta_{{\cal O}_{\phi^0}} = 4, \hspace{0.8 in} \Delta_{{\cal O}_{\phi^r}} = 3\,\,,\,\,\,\,r=1,2,3
\eea

In \cite{Ferrara:1998gv} these CFT operators were expressed in terms of free hypermultiplets (i.e. the singleton sector). The case of $n=1$ corresponds to having a single free hypermultiplet, consisting of four real scalars $q_A^I$ and two symplectic Majorana spinors $\psi ^I$. Here $I=1,2$ is the $SU(2)_R$ R-symmetry index and $A=1,2$ is the $SU(2)$ flavor symmetry index. The gauge invariant operators appearing in \eqref{opedic} are related to these fundamental fields as follows,
\bea
\label{massdeformations}
{\cal O}_{\sigma}=(q^*)^A_{\;\;I} q_{\;\;A}^I,  \hspace{0.4 in} {\cal O}_{\phi^0}= \bar \psi_I \psi^I,  \hspace{0.4 in}{\cal O}_{\phi^r}=  (q^*) ^A_{\;\;I}  (\sigma^r)_A^{\;\; B}  {q^I}_B\,\,, \,\,\,\,\,r=1,2,3
\eea
Note that the first two operators correspond to mass terms for the scalars and fermions, respectively, in the hypermultiplet. The third operator is a triplet with respect to the $SU(2)_R$ R-symmetry. 
As argued in \cite{Ferrara:1998gv}, the field $\phi^0$ is the top component of the global current supermultiplet. Therefore a deformation by ${\cal O}_{\phi^0}$ will break superconformal symmetry but preserve all Poincare supersymmetry \cite{Cordova:2016xhm}. However, deformation by ${\cal O}_{\phi^0}$ alone is inconsistent. Poincare supersymmetry demands that we also turn on the scalar masses ${\cal O}_{\sigma}$. Moreover, supersymmetry on $S^5$ requires an additional operator in the action that breaks the superconformal $SU(2)_R$ symmetry to $U(1)_R$ symmetry \cite{Hosomichi:2012ek}. Without loss of generality, we may choose this operator to be ${\cal O}_{\phi^3}$.

%%%%%%%%%%%%%%%%%%%%%%%%%%%%%%%%%%%%%%%%%%%%
%%%%%%%%%%%%%%%%%%%%%%%%%%%%%%%%%%%%%%%%%%%%
\section{Euclidean theory and BPS solutions}
\setcounter{equation}{0}
\label{sec3}
%%%%%%%%%%%%%%%%%%%%%%%%%%%%%%%%%%%%%%%%%%%%
%%%%%%%%%%%%%%%%%%%%%%%%%%%%%%%%%%%%%%%%%%%%

In this section we will obtain the six-dimensional holographic dual of a mass deformation of a 5D SCFT on $S^5$. Such a dual is given by $S^5$-sliced domain wall solutions of matter-coupled Euclidean $F(4)$ gauged supergravity. In order to obtain such solutions, we must first continue the Lorentzian signature gauged supergravity outlined above to Euclidean signature, which has subtleties for both the scalar and fermionic sectors. Once the Euclidean theory is obtained, we turn on relevant scalars necessary to support the domain wall. As discussed in the previous section, at least three scalars must be turned on to obtain supersymmetric solutions. The ansatz for the domain wall solutions takes the following form
\bea
\label{dwansatz}
ds^2 = du^2 + e^{2f(u)} ds_{S^5}^2,\hspace{0.4in} \sigma=\sigma(u),\hspace{0.4in}  \phi^i=\phi^i(u), \hspace{0.1 in}i=0 ,3
\eea
with the remaining fields set to zero. Next we will obtain a consistent set of BPS equations on  the above ansatz, and then solve them numerically. When solving them, we will demand as an initial condition that for some finite $u$ the metric factor $e^{2f}$ vanishes, so that the geometry closes off smoothly.

\subsection{Euclidean action}

The Euclidean action may be obtained from the Lorentzian one by first performing a simple Wick rotation of Lorentzian time $t\rightarrow-ix^6$. This makes the spacetime metric negative definite, since the metric in the Lorentzian theory was taken to be of mostly negative signature. However, we will choose to work with the Euclidean theory with positive definite metric. Making this modification involves a change in the sign of the Ricci scalar. Then noting that the Euclidean action is related to the Lorentzian action by $\exp \(i S^{Lor}\)=\exp \(-S^{Euc}\)$, the final result of the Wick rotation is the following Euclidean action,
\be
\label{Euclideanaction}
S_{6D}=\frac{1}{4\pi G_6}\int d^6x ~\sqrt{G} \cL ~,~~~~ \cL= \(-{1 \over 4} R +\p_\m\sigma\p^\m\sigma+{1\over 4}G_{ij}(\phi)\p_\m \phi^i \p^\m \phi^j + V(\sigma,\phi^i)\)
\ee
where the spacetime metric $G$ is positive definite and $G_6$ is the six-dimensional Newton's constant. By abuse of notation, $G_{ij}(\phi)$ with indices refers to the metric on the scalar manifold, which for the coset representative \eqref{cosetrep} is given by
\be
G_{ij}=\text{diag}\(\cosh^2 \phi^1 \cosh^2 \phi^2 \cosh^2 \phi^3, \cosh^2 \phi^2 \cosh^2 \phi^3, \cosh ^2 \phi^3, 1\)
\ee

In addition to performing the above Wick rotation, we also perform a Wick rotation on the sigma model \cite{Bergshoeff:2008be,Hertog:2017owm,Ruggeri:2017grz} 
\bea
{SO(4,1) \over SO(4)} \to{SO(4,1) \over SO(3,1)} \simeq dS_4
\eea
The metric on the sigma model is now that of dS$_4$, as opposed to the $\mathbb{H}_4$ that we had in the Lorentzian case \cite{Gutperle:2017nwo}. This can be obtained by making the following change to the $\mathbb{H}_4$ coset,
\bea
\label{cosetcont}
 \phi_r\rightarrow i \phi_r \hspace{0.5 in}r=1,2,3
 \eea
It would be interesting to understand this analytic continuation from first principles and its relation to Euclidean supersymmetry, possibly along the lines of \cite{Gibbons:1995vg,Cortes:2003zd}. For now, we just note that such a Wick rotated model seems necessary to obtain regular, supersymmetric solutions. 

\subsection{Euclidean supersymmetry}

The next task is to identify the form of the Euclidean supersymmetry variations. Motivation for the form of these variations may be obtained by analysis of the free differential algebra (FDA) of the $F(4)$ gauged supergravity theory with $\mathbb{H}_6$ vacuum, as discussed in Appendix \ref{cosrep}. The final result for this FDA is given in (\ref{H6FDA}), and is noted to be of the same form as the FDA for the theory with dS$_6$ background (identified in \cite{DAuria:2002xnk}), with two differences. The first obvious difference is that the metrics differ - the space considered in \cite{DAuria:2002xnk} was dS$_6$ with mostly minus signature, whereas we are currently focused on positive definite $\mathbb{H}_6$. However, both of these spaces have $R_{\mu\nu} = - 20 m^2 g_{\mu\nu}$. The second difference is in the definition of Dirac conjugate spinors. However, once the difference in definition of the gamma matrices is accounted for, the only difference is a factor of $i$, i.e.
\bea
\bar \psi_A^{(\mathbb{H}_6)} =  i \bar \psi_A^{(dS_6)} 
\eea
Because of these similarities, the supersymmetry variations in the current case are expected to be of a similar form to that of \cite{DAuria:2002xnk}. In particular, the variations of the fermions are expected to be of the form
\bea
\label{susyvariations}
\delta \chi_A &=& - \half \g^\m \p_\m \sigma \ep_A + N_{AB} \ep^B + \dots
\no\\\no\\
\delta \psi_{A \m} &=& \cD_\m \ep_A + i S_{AB} \g_\m \ep^B +\dots
\no\\\no\\
\delta \lambda^I_A &=& - \hat P^I_{r i} \sigma^r_{AB} \p_\m \phi^i \g^\m \ep^B + \hat P^I_{0 i} \epsilon_{AB} \p_\m \phi^i \g^7 \g^\m \ep^B + M^I_{AB} \ep^B+\dots
\eea
where $N_{AB}$, $S_{AB}$, and $M^I_{AB}$ are again given by (\ref{SMNvars}), but now with the appropriate redefinition of the coset representative as per (\ref{cosetcont}). It should be noted that while the FDA analysis presented in Appendix \ref{cosrep} is a strong motivation for the form of the supersymmetry variations presented above, it is not a proof. To actually derive the form of these variations, one must first introduce curvature terms representing deviations from zero of each line in the free differential algebra. An application of the exterior derivative to the resulting expressions then gives rise to Bianchi identities, which must be solved before obtaining the explicit form of the fermion variations. This is a rather involved process, and so for the moment we will content ourselves with the motivating comments provided by the FDA. We will take the eventual presence of smooth supersymmetric solutions consistent with the equations of motion as \textit{a posteriori} evidence for the legitimacy of these variations.

A nice property of the variations above is the fact that they are consistent with the following $SO(6)$-invariant symplectic Majorana condition, 
\bea
\bar \psi_A = \epsilon^{AB} \psi_B^T \cC
\eea
The consistency of such a condition allows us to work with symplectic Majorana spinors just as in the Lorentzian case, though the symplectic Majorana condition utilized here is different than that of the Lorentzian case.\footnote{The fact that the symplectic Majorana condition must be different in the current case follows from $SO(6)$ invariance. The condition used in the Lorentzian case \cite{Gutperle:2017nwo} was expressed in terms of $\g_0$, which explicitly breaks $SO(6)$ symmetry. }

As mentioned before, we will be concerned with only the simplest case of a single non-zero $SU(2)_R$-charged vector multiplet scalar $\phi^3$, i.e. we take $\phi^1=\phi^2=0$. It can be easily verified that this is a
consistent truncation, and is in fact the most general choice of non-vanishing fields that can preserve $SO(4, 2) \times U(1)_R$. With this consistent truncation, the functions $N_{AB}$, $S_{AB}$, and $M^I_{AB}$ appearing in the supersymmetry variations reduce to 
\bea
S_{AB} &=& i S_0 \epsilon_{AB} + i S_3 \g^7 \sigma^3_{AB}
\no\\
N_{AB}&=& -N_0 \epsilon_{AB} -N_3 \g^7 \sigma^3_{AB}
\no\\
M^I_{AB}&=& M_0 \g^7 \epsilon_{AB} + M_3  \sigma^3_{AB}
\eea
where we have defined 
\bea
\label{newdefs}
S_0&=&\frac14 \left(g\cos \phi^3 e^\s+m e^{-3\s}\cosh \phi^0\right)
\no\\
S_3&=&\frac14 i \,m ~e^{-3\s}\sinh \phi^0 \sin \phi^3
\no\\
N_0&=&-\frac14 \left(g\cos \phi^3 e^\s-3m e^{-3\s}\cosh \phi^0\right)
\no\\
N_3&=&-\frac34 i\, m e^{-3\s}\sinh \phi^0 \sin \phi^3
\no\\
M_0&=&2m ~e^{-3\s}\cos \phi^3\sinh \phi^0
\no\\
M_3&=&-2 i\, g ~e^{\s}\sin \phi^3
\eea
Importantly, note that $S_3$, $N_3$, and $M_3$ are now purely imaginary, in contrast to the Lorentzian case \cite{Gutperle:2017nwo}. In all that follows we will set $m=-1/2\, \eta$ such that the radius of AdS$_6$ is one.

\subsection{BPS Equations}

We now use the vanishing of the fermionic variations (\ref{susyvariations}) to obtain BPS equations for the warp factor and the three non-zero scalars. 
\subsubsection{Dilatino equation and projector}
We begin by imposing the vanishing of the dilatino variation, $\delta\chi_A = 0$, which implies
\bea
\label{dilatinoeq}
\half \g^5 \sigma' \ep_A = N_0 \ep_A + N_3 \g^7 {(\sigma^3)^B}_A \ep_B
\eea
This equation can be interpreted as a projection condition on the spinors $\ep_A$. Consistency of this projection condition then requires that 
\bea
\label{sigmaBPSeq}
 \sigma' = 2 \eta \sqrt{N_0^2 + N_3^2}
\eea
where $\eta = \pm1$. Plugging this BPS equation back into (\ref{dilatinoeq}) then yields a second form of the projection condition,
\bea
\g^5 \ep_A = G_0 \ep_A - G_3 \g^7 {(\sigma^3)^B}_A \ep_B
\eea
which is more useful in the derivation of the other BPS equations. In the above, we have defined 
\bea
\label{G03defs}
G_0 = \eta {N_0 \over \sqrt{N_0^2 + N_3^2}}\hspace{1 in} G_3 = -\eta {N_3 \over \sqrt{N_0^2 + N_3^2}}
\eea
\subsubsection{Gravitino equation}
The analysis of the gravitino equation  $\delta \psi_{A \m}=0$ proceeds in exactly the same way as for the Lorentzian case studied in \cite{Gutperle:2017nwo}. The procedure gives rise to a first-order equation for the warp factor $f$ and an algebraic constraint. To avoid excessive overlap with that paper, we simply cite the result,
\bea
f' = 2 (G_0 S_0 + G_3 S_3) \hspace{0.5 in}
e^{-2f} = 4 (G_0 S_0 + G_3 S_3)^2 - 4 (S_0^2 + S_3^2)
\eea
\subsubsection{Gaugino equations}
Finally, we turn toward the gaugino equation $\delta \lambda^I_A=0$. Again the analysis of this equation proceeds in an exactly analogous manner to the Lorentzian case \cite{Gutperle:2017nwo}. The result is
\bea
\cos \phi^3 (\phi^0)' = - (G_0 M_0 + G_3 M_3) \hspace{0.5 in} (\phi^3)' = i (G_3 M_0 - G_0 M_3)
\eea
The right-hand sides of both equations are real, and thus give rise to real solutions when appropriate initial conditions are imposed.
%%%%
\subsubsection{Summary of first-order equations}
%%%%
To summarize, the first-order equations for the warp factor $f$ and the scalars $\s,\phi^0,\phi^3$ are found to be
\bea
f'&=&2\(G_0S_0+ G_3S_3\)
\nonumber \\%\label{BPSeq2}
 \s'&=&2 \eta \sqrt{N_0^2 + N_3^2}
\nonumber \\%\label{BPSeq3}
\cos \phi^3 \(\phi^0\)'&=&-\(G_0M_0+G_3M_3\)
\nonumber\\%\label{BPSeq4}
 \(\phi^3\)'&=&i \(G_3 M_0-G_0M_3\)
 \label{BPSeq1}
\eea
Furthermore, for consistency these were required to satisfy the algebraic constraint
\be\label{constrainta}
e^{-2f}=4\(G_0S_0+G_3S_3\)^2-4 \(S_0^2+S_3^2\)
\ee
The various functions featured in these equations were defined in (\ref{newdefs}) and (\ref{G03defs}).

%%%%
\subsection{Numeric solutions}
In order to get acceptable numerical solutions from these equations, we must choose appropriate initial conditions. It is easy to check that the following initial conditions ensure smoothness of all three scalars, as well as the vanishing of $e^{2f}$ at the origin,
\bea
\label{initialconds}
 \phi^3_0 &=& \sin^{-1} \left[{1 \over 8 \tanh\phi^0_0}\left(-3 + \sqrt{9 + 16 \tanh^2 \phi^0_0}\right) \right]
\no\\
\sigma_0 &=& {1 \over 4} \log \left[{\cosh \phi^0_0 \left(5 + \sqrt{9 + 16 \tanh^2 \phi^0_0} \right) \over \sqrt{6}\sqrt{8 + \coth^2 \phi^0_0 \left(-3 + \sqrt{9 + 16 \tanh^2 \phi^0_0} \right)}}\right]
\eea
We have defined for notational convenience $\phi^\a_0 \equiv \phi^\a(0)$ and $\sigma_0 \equiv \sigma(0)$. For these initial conditions to be real, we must ensure that
\bea
|f(\phi^0_0)| \leq 1 \hspace{0.7 in} f(\phi^0_0) \equiv {1 \over 8 \tanh\phi^0_0}\left(-3 + \sqrt{9 + 16 \tanh^2 \phi^0_0}\right) 
\eea
Noting that 
\bea
\lim_{\phi^0_0 \rightarrow - \infty} f(\phi^0_0) = - {1 \over 4} \hspace{0.8 in} \lim_{\phi^0_0 \rightarrow + \infty} f(\phi^0_0) ={1 \over 4}
\eea
and also that $f(\phi^0_0)$ is monotonically increasing, i.e.
\bea
{d f \over d \phi^0_0} >0\hspace{0.3in} \forall \phi^0_0 \in \RR
\eea
allows us to conclude that this is always the case for real initial conditions $\phi^0_0$. Thus we have a one parameter family of real smooth solutions, labeled by the IR parameter $\phi^0_0$.

{
\begin{figure}[h]
\centering
\begin{minipage}{.5 \textwidth} 
\centering
\includegraphics[scale=0.7]{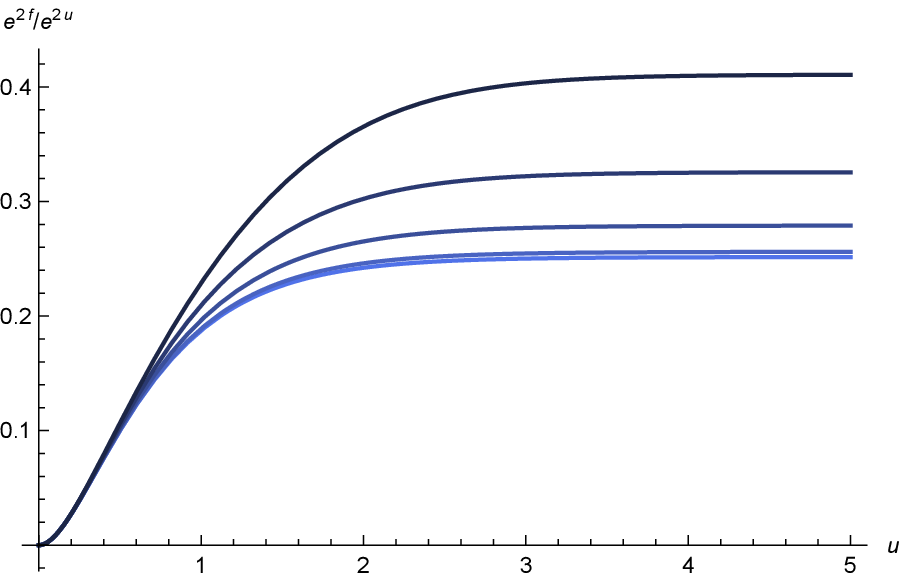}
\end{minipage}%
\begin{minipage}{.5 \textwidth} 
\centering
\includegraphics[scale=0.7]{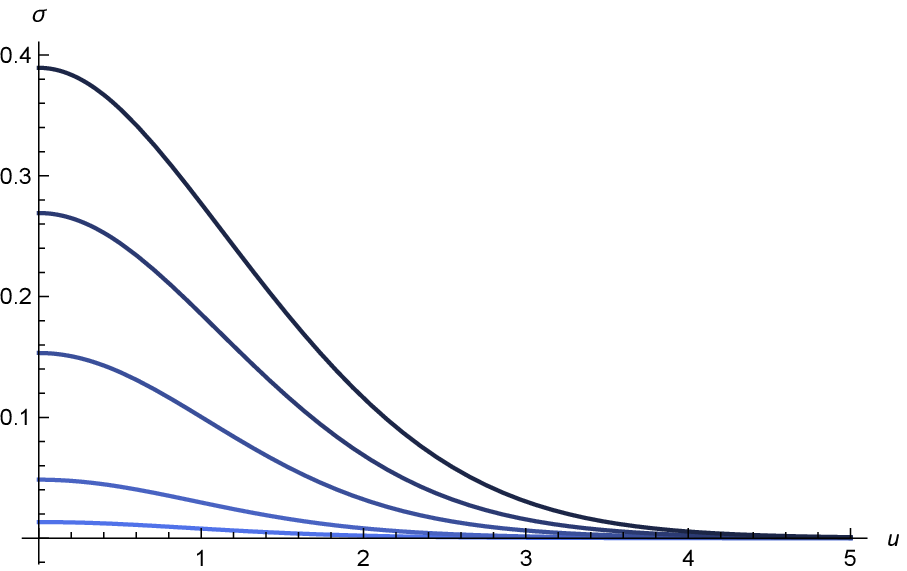}
\end{minipage}%
\newline
\newline
\newline
\centering
\begin{minipage}{.5 \textwidth} 
\centering
\includegraphics[scale=0.7]{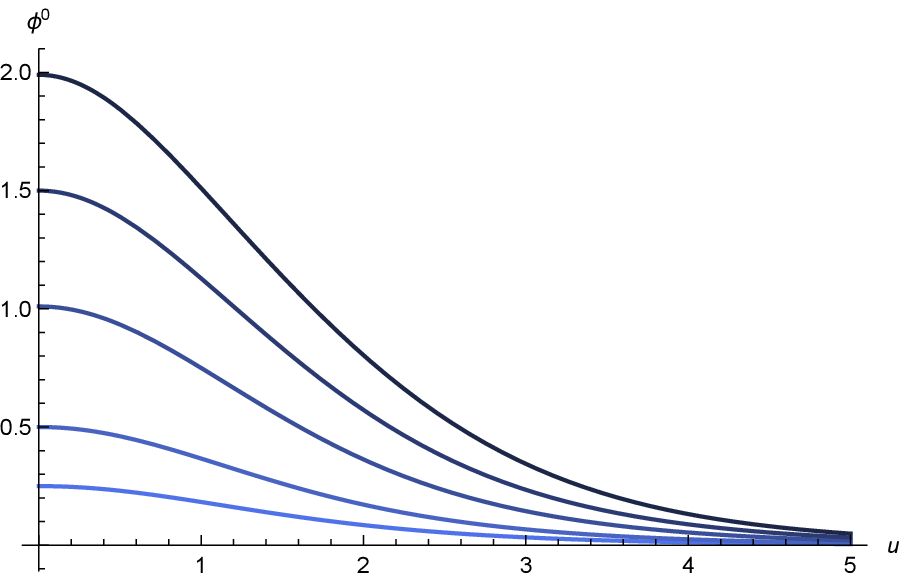}
\end{minipage}%
\begin{minipage}{.5 \textwidth} 
\centering
\includegraphics[scale=0.7]{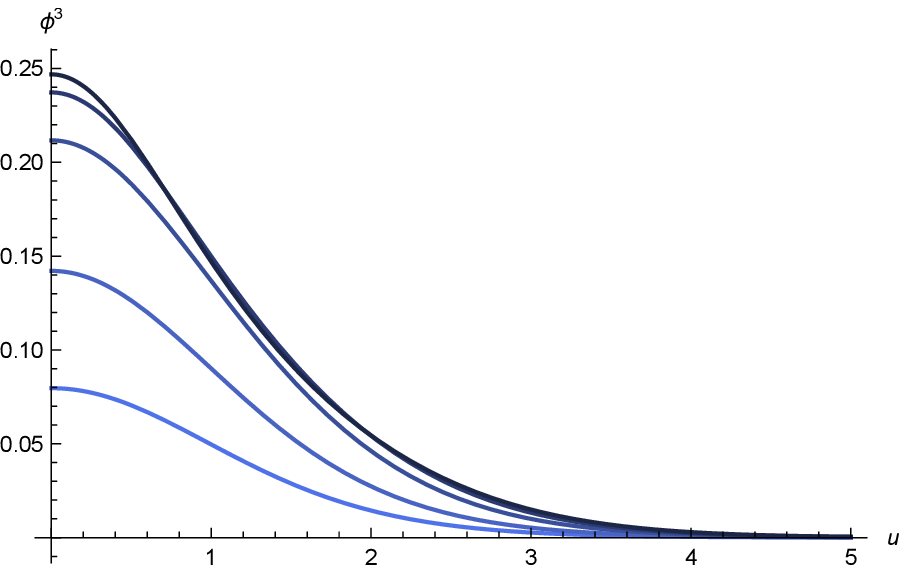}
\end{minipage}%
\caption{Smooth solutions for the four scalar fields in the Euclidean theory. We take $\eta=-1$ and have chosen the following values for the initial conditions: $\phi^0_0=\{0.25,\,0.5,\,1,\,1.5,\,2 \}$ (light to dark blue). Importantly, we see that $e^{2f}$ vanishes at the origin - signaling a smooth closing off of the spacetime - and asymptotes to a constant $e^{2 f_k}$.}
\label{fig1}
\end{figure}
}
With this in mind, we may choose any value of $\phi^0_0$ and solve the BPS equations in (\ref{BPSeq1}) numerically. In Figure \ref{fig1}, we plot the solutions obtained for the following choices of initial condition: $\phi^0_0=\{0.25,\,0.5,\,1,\,1.5,\,2 \}$. In order to get smooth solutions for $u>0$, we must take $\eta = -1$. It is straighforward to verify that  the resulting solutions are completely smooth and have the expected vanishing of $e^{2f}$ at the origin, implying that the spacetime smoothly pinches off. Furthermore, $e^{2f}/e^{2u}$ is seen to asymptote to a constant, which we denote by $e^{2 f_k}$.

\subsection{UV asymptotic expansions}

As in the holographic Janus solutions in Lorentzian signature \cite{Gutperle:2017nwo}, the BPS equations may also be used to obtain the UV asymptotic behavior of the solutions. To do so, we begin by defining an asymptotic coordinate $z = e^{- u}$, where the asymptotic $S^5$  boundary is reached by taking $u\to \infty$. Consequently, an asymptotic expansion is an expansion around $z=0$. The coefficients in the UV expansions of the non-zero fields may now be solved for order-by-order using the BPS equations. One finds explicitly that all coefficients are determined in terms of only three independent parameters $\a$, $\b$, and $f_k$, in accord with the fact that there are three independent first-order differential equations. The first few terms in the expansions are
\bea
\label{UVexp}
f(z) &=& -\log z +f_k -  \left({1\over 4}e^{-2f_k} + {1\over 16}\a^2\right)  \,z^2 + O(z^4) \no\\
\sigma(z) &=& {3\over 8} \a^2 \,z^2 +  {1\over 4} e^{f_k} \a \b \,z^3 + O(z^4) 
\no\\
\phi^0(z) &=& \a \, z - \left({5\over 4}\a \,e^{-2f_k} + {23\over 48} \a^3\right) \,z^3 + O(z^4)
\no\\
\phi^3(z) &=& e^{-f_k} \a  z^2 + \b \, z^3 + O(z^4) 
\eea
We have obtained the expansions up to $O(z^8)$, but we display only the first few terms here.

%%%%%%%%%%%%%%%%%%%%%%%%%%%%%%%%%%%%%%%%%%%%
%%%%%%%%%%%%%%%%%%%%%%%%%%%%%%%%%%%%%%%%%%%%
\section{Holographic sphere free energy}
\setcounter{equation}{0}
\label{sec4}
%%%%%%%%%%%%%%%%%%%%%%%%%%%%%%%%%%%%%%%%%%%%
%%%%%%%%%%%%%%%%%%%%%%%%%%%%%%%%%%%%%%%%%%%%

The goal of this section is to obtain the holographic free energy, i.e. the renormalized on-shell action. We begin by writing the full action, 
\bea
\label{introact}
&\vphantom{.}& \hspace{1.7 in} S = S_{\tx{6D}} + S_{\tx{GH}}
\no\\\no\\
&\vphantom{.}& S_{\tx{6D}} = \int du~d^5 x ~\sqrt{G}\, \cL  \hspace{1 in}S_{\tx{GH}}=- {1 \over 2} \int d^5 x \sqrt{\g}\, \cK
\eea
where $S_{\tx{6D}}$ is the six-dimensional Euclidean action given in (\ref{Euclideanaction}) and $S_{\tx{GH}}$ is the Gibbons-Hawking term.\footnote{We have set $4\pi G_6=1$ to avoid clutter in the formulas. We will restore this factor in the final expression for the free energy.} The $\g$ appearing in $S_{\tx{GH}}$ is the determinant of the induced metric on the boundary (located at some cutoff distance $u= \Lambda$), while $\cK$ is the trace of the extrinsic curvature $\cK_{ij}$ of the radial $S^5$ slices. The latter is defined as
\be
\cK_{ij}=\frac12 \frac{d}{du}\gamma_{ij} 
\ee

In general, the on-shell action is divergent and requires renormalization. The addition of infinite counterterms is standard in holographic renormalization   \cite{Bianchi:2001kw,Skenderis:2002wp,Papadimitriou:2004rz}, but in the current case we must also add finite counterterms in order to preserve supersymmetry \cite{Bianchi:2001de}.  We will begin our exploration of counterterms in this section by first considering the finite counterterms in the limit of a flat domain wall, after which we move onto infinite counterterms in the more general case of a curved domain wall. Finally, appropriate curved space finite counterterms will be fixed by demanding finiteness of the one-point functions of the dual operators.

\subsection{Finite counterterms}
In order to obtain finite counterterms, we will make use of the Bogomolnyi trick \cite{Bobev:2013cja,Bobev:2016nua,Freedman:2013ryh}. To do so, we will first need to identify a superpotential $W$. Though we will find that no exact superpotential can be found for our solutions - in the sense that there is no superpotential which can recast all of the BPS equations in gradient flow form -  we will be able to identify an \textit{approximate} superpotential. By ``approximate" here, we mean that it does yield gradient flow equations up to terms of order $O(z^5)$, where the asymptotic coordinate $z$ was defined earlier as $z=e^{-u}$. This is useful since, as we will see later, we will only need terms up to $O(z^5)$ to obtain all divergent and finite counterterms. Terms of higher order will all vanish in the $\epsilon \rightarrow 0$ limit, i.e. when the UV cutoff is removed. Thus the approximate superpotential will yield  all finite counterterms. 

\subsubsection{Approximate superpotential}
In order to identify a candidate superpotential, we begin by recalling the form of the scalar potential $V$. With the choice of coset representative and consistent truncation outlined in Section \ref{sec3}, one finds that
\bea
V(\sigma,\phi^i) = -9 m^2 e^{2 \sigma }-12m^2 e^{- 2 \sigma} \cosh \phi^0 \cos \phi^3 + m^2 e^{-6 \sigma} \cosh^2 \phi^0+m^2 e^{-6 \sigma} \cos 2 \phi^3 \sinh^2 \phi^0
\no
\eea
 This scalar potential can in fact be rewritten as
\bea
V = 4 (N_0^2 + N_3^2) + {1 \over 4} (M_0^2 +M_3^2) - 20 (S_0^2 + S_3^2)
\eea
Then for BPS solutions, (\ref{BPSeq1}) implies that 
\bea
\label{VBPS}
V = (\sigma')^2 + {1 \over 4}\left(-({\phi^3}')^2 + \cos^2 \phi^3 ({\phi^0}')^2\right)- 20 (S_0^2 + S_3^2)
\eea
This motivates us to define a superpotential $W$ as 
\bea
\label{superpot}
W = \sqrt{S_0^2 + S_3^2}
\eea
\iffalse
This can also be motivated from the integrability of the gravitino equation. Namely, one would usually expect integrability to give something proportional to $W^2$, i.e.
\bea
[D_m , D_n] \eps_A = 2 W^2 \,\g_{mn} \eps_A
\eea
and indeed one finds (\ref{integrability1}).

Requiring that the warp factor BPS equation may be re-expressed as an equation linear in the superpotential (\ref{superpot}), i.e.
\bea
\label{fgradflow}
f' &=& 2 (G_0 S_0 + G_3 S_3)
\no\\
&=& 2 \, \g \,W
\eea
gives us a factor $\g$ defined as
\bea
\g = {G_0 S_0 + G_3 S_3 \over \sqrt{S_0^2 + S_3^2}}
\eea
We now make the non-trivial check that, with these definitions of $W$ and $\g$, the BPS equation for $\s$ takes the expected gradient flow form, 
\bea
\label{sigmagradflow}
\s ' &=&2 \eta \, \sqrt{N_0^2 + N_3^2}
\no\\
&=& -2 \left({- S_0 N_0 + S_3 N_3 \over G_0 S_0 + G_3 S_3}\right)
\no\\
&=& -2\, \g^{-1} \,\p_\s W
\eea
exactly as expected.
\fi
Unfortunately, this superpotential does \textit{not} allow one to write the BPS equations for both $\phi^0$ and $\phi^3$ as gradient flow equations. The reason for this failure is that the integrability condition required to convert the BPS equation into a gradient flow form is not satisfied; see e.g. Appendix C.2.1 of \cite{Bobev:2013cja}.\footnote{See however \cite{Lindgren:2015lia,Cabo-Bizet:2017xdr} where an effective superpotential involving the warp factor was derived, in terms of which the first-order equations take the form of a gradient flow.} We thus follow the strategy of \cite{Bobev:2013cja} to construct an approximate superpotential. Our model consists of two consistent truncations that admit flat domain walls and an exact superpotential. These are the $\phi^3=0,\phi^0\ne0$ truncation and the $\phi^0=0,\phi^3\ne0$ truncation. The corresponding flow equations are (we set $\eta=-1$ henceforth)
\bea
\label{x0gradflow}
{\phi^0}' = -8 \, \p_{\phi^0} W|_{\phi^3=0} \hspace{1 in} {\phi^3}' =8 \, \p_{\phi^3} W|_{\phi^0=0}
\eea
respectively. In either truncation, the BPS equations for the warp factor and dilaton $\s$ can be put in the following form,
\be\label{warpflat}
f'=2\,W\hspace{1 in} \s'=2 \,\partial_\s W
\ee
An important fact is that, though the gradient flow equations of (\ref{x0gradflow}) do not hold exactly in the full model with $\phi^0\ne0,\phi^3\ne0$, they \textit{do} hold up to and including $O(z^5)$. Looking at the form of the UV asymptotics of the scalar fields, one may expand the superpotential of (\ref{superpot}) keeping only terms contributing up to this order. This gives 
\bea
\label{approxsuperpot}
W =  \half + {3 \over 4} \sigma^2 + {1 \over 16} (\phi^0)^2 -{3 \over 16} (\phi^3)^2 + {1 \over 192} (\phi^0)^4 -{3 \over 16} (\phi^0)^2 \sigma + \dots
\eea
where the dots represent terms of order $O(z^6)$. This is the approximate superpotential we will use in what follows.

\subsubsection{Bogomolnyi trick}
We now use the Bogomolnyi trick \cite{Bobev:2013cja,Bobev:2016nua,Freedman:2013ryh} to get the finite counterterms needed to preserve supersymmetry in the case of a flat domain wall. The central idea of the Bogomolnyi trick is that for a BPS solution, the renormalized on-shell action must vanish. In order to make use of this fact, we will first want to recast the on-shell action in a simpler form.

To do so, we begin by inserting (\ref{VBPS}) into (\ref{Lagrangian}). We find that 
\bea
\cL =- {1 \over 4}R -20 W^2 + 2 \cL_{\tx{kin}}
\eea
where we've defined 
\bea
\cL_{\tx{kin}} = (\sigma')^2 + {1 \over 4}\left[-({\phi^3}')^2 + \cos^2 \phi^3 ({\phi^0}')^2\right]
\eea
The non-zero components of the Ricci tensor are
\bea
\label{RicciTens}
R_{uu} = - 5 \left(f'' + (f')^2 \right) \hspace{1 in} R_{mn} =  -g_{mn} \left(f'' + 5(f')^2 \right)
\eea
while the Ricci scalar is given by
\bea
R = - 10 f'' - 30(f')^2 
\eea
Furthermore, we have that $\sqrt{G} = e^{5f} \sqrt{g}$, where $g$ is the determinant of the unit $S^5$ metric. Upon integration by parts, part of the Einstein-Hilbert term cancels with the Gibbons-Hawking term to give the following simple expression
\bea
\label{action1}
S = \int du \int d^5 x \sqrt{g}\, e^{5f}\left[-5 \left((f')^2 +4 W^2 \right)+2 \cL_{\tx{kin}}\right]
\eea
The restriction to the flat case was not strictly necessary so far, but it will be crucial in the next step.
The gradient flow equations \eqref{x0gradflow} and \eqref{warpflat}, together with the chain-rule, allows us to rewrite
\bea
\cL_{\tx{kin}} =-2\,  W'
\eea
Plugging this into (\ref{action1}) and using the BPS equation of the warp factor, we find
\bea
\label{SLam}
S = -4  \int d^5 x \sqrt{g}\,e^{5f} W \Big|_0^\Lambda
\eea
where $\Lambda$ is the UV cutoff. Only the $\Lambda$ part of the action contributes, since $e^{5f} W|_0$ vanishes due to the close-off of the geometry. 

Removing the UV cutoff $\Lambda\rightarrow \infty$ is equivalent to removing the cutoff $\ep$ on our asymptotic coordinate $z$, i.e. $\ep\rightarrow 0$. From the UV asymptotics (\ref{UVexp}) we find that in this limit the factor $e^{5f}$ diverges like 
\bea
e^{5f} \sim {1 \over \ep^5}
\eea
This is the reason for the previous claims that only the terms up to $O(z^5)$ in the superpotential are relevant for obtaining counterterms. All the higher-order terms vanish as the cutoff is removed. 
We may thus legitimately insert the approximate superpotential (\ref{approxsuperpot}) into (\ref{SLam}) to get the counterterms, 
\bea
\label{SLam2}
S^{(W)}_{\tx{ct}} = 4 \int d^5x \sqrt{\g} \left[  \half + {3 \over 4} \sigma^2 + {1 \over 16} (\phi^0)^2 -{3 \over 16} (\phi^3)^2 + {1 \over 192} (\phi^0)^4 -{3 \over 16} (\phi^0)^2 \sigma \right]
\eea
where $\g$ is the induced metric on the $z = \ep$ boundary. All fields are evaluated at $z = \ep$. This gives all finite and infinite counterterms for the flat domain wall solutions.

\subsection{Infinite counterterms}
We now turn towards the identification of the infinite counterterms in the more general curved domain wall case. We may first solve for all of the infinite counterterms via the usual holographic renormalization procedure. Once we have these, we will
\begin{enumerate}
\item Check that in the flat limit, they reduce to the divergent pieces of the flat counterterms (\ref{SLam2}) found above.
\item Add to them the finite pieces found in (\ref{SLam2}) but missing in the holographic renormalization procedure. 
\end{enumerate}
For simplicity, we will perform holographic renormalization on supersymmetric solutions only, and thus the infinite counterterms we obtain are universal for supersymmetric solutions only. 

We begin by using the expression for the on-shell Ricci scalar, 
\bea
R = 4 (\sigma')^2 + \left[-({\phi^3}')^2 + \cos^2 \phi^3 ({\phi^0}')^2 \right] + 6 V
\eea
to rewrite the action (\ref{introact}) as 
\bea
S_{\tx{6D}} &=& -{1 \over 2} \int du\, d^5 x \sqrt{g} \,e^{5f}  V
\eea
We have not included the Gibbons-Hawking term yet, but will do so later. The first step of holographic renormalization is to isolate the divergent terms. We may do so by expanding all fields using their UV asymptotics, then integrating over small $z$ and evaluating on the cutoff $\eps$. Doing so, we find 
\bea
S_{\tx{6D}} &=& -\half \int d^5 x \sqrt{g} e^{5 f_k} \left[{1 \over \eps^5} + {1 \over 3 \eps^3}\left(25 f_2 + \(\phi^0_1\)^2 \right) \right.
\no\\
&&\hspace{1.3 in} + {1 \over 24\eps}\left(1500 f_2^2 + 600 f_4 + 120 f_2 \(\phi^0_1\)^2  - \(\phi^0_1\)^4 \right.\no\\
&\vphantom{.}& \hspace{1.1 in}\left. \left.  \hspace{1.3 in}+ 48\, \phi^0_1 \phi^0_3 + 36\left(-\(\phi^3_2\)^2 + 4 \sigma_2^2\right)  \right)\right]
\eea
where we've thrown out all non-divergent contributions. Note that the integration would naively give a $\log \eps$, but this vanishes on the BPS equations since they constrain the UV asymptotic expansion coefficients in the following way,\footnote{We have shown this using the solutions of the BPS equations, but it must hold for general solutions of the equations of motion as well.} 
\bea
25 f_5 + 2 \,\phi^0_1 \phi^0_4 - 3 \,\phi^3_2 \phi^3_3 + 12 \sigma_2 \sigma_3 = 0
\eea
The absence of the logarithmic term is to be expected, since any dual five-dimensional field theory is anomaly-free. The Gibbons-Hawking term is 
\bea
S_{\tx{GH}} = -{5 \over 2} \int d^5 x \sqrt{g}\, e^{5f} f'
\eea
We again use the asymptotic expansions to write 
\bea
S_{\tx{GH}} = -{5 \over 2} \int d^5 x \sqrt{g} e^{5 f_k} \left[{1 \over \eps^5} +{3 f_2\over \eps^3} +{1 \over2 \eps} \left(5 f_2^2 + 2 f_4\right) \right]
\eea
Adding the two together, we find in total that 
\bea
\label{Scurvcount}
S_{\tx{6D}}+S_{\tx{GH}} &=& -\int d^5 x \sqrt{g} e^{5 f_k} \left[{2 \over \eps^5} +{1 \over 6 \eps^3}\left(20 f_2 - \(\phi^0_1\)^2 \right) \right.- {1 \over 48\eps}\left(1200 f_2^2 + 480 f_4  \right.  \no\\
&\vphantom{.}&\hspace{0.2 in}\left. \left. \hspace{.5in} + 120 f_2 \(\phi^0_1\)^2 - \(\phi^0_1\)^4 + 48\, \phi^0_1 \phi^0_3 -36(\phi^3_2)^2 + 144 \sigma_2^2  \right)\right]
\eea
We must now undergo the task of inverting all of the UV modes to rewrite the action in terms of induced fields at the cut-off surface (since it is the latter which transform nicely under bulk diffeomorphism). Before quoting the result, we note that at the cut-off $z = \eps$, the induced metric $\g_{i j}$ is given by 
\bea
\g_{i j}=  e^{2 f}\big|_{z = \eps} \,g^{(S^5)}_{i j}
\eea  
The Ricci tensor and Ricci scalar are given by
\bea
R_{ij}[\gamma] = 4e^{-2f}\gamma_{ij}\big|_{z = \eps}~~~~~~~~~~~~~~R[\gamma] = 20 \,e^{- 2 f}\big|_{z = \eps}
\eea
In terms of these quantities, we find that the inverted form of the divergent part of the on-shell action is 
\bea
\label{totSdiv}
S &=& - \int d^5 x \sqrt{\g} \left[2 + {1 \over 4} \left(\phi^0\right)^2 + {3 \over 4} \left(\phi^3\right)^2 - 3 \sigma^2 +{7 \over 12 }\left(\phi^0\right)^4\right.
\no\\
&\vphantom{.}&\hspace{1.4 in}\left. + {1 \over 12} R[\g] - {1 \over 320} R[\g]^2 - {3 \over 32} R[\g] \(\phi^0\)^2 \right]
\eea
We may now address the two points mentioned at the start of this subsection. To begin, we check that in the flat limit, we reproduce the divergent terms obtained in (\ref{SLam2}). In particular, we expect that the first line of (\ref{totSdiv}) should be equal to $-S_{ct}^{(W)}$ up to and including order $O(z^4)$. Though the expressions look different at first sight, it can be checked via the relationships between expansion coefficients in (\ref{UVexp}) (along with their higher order counterparts) that in the limit $e^{-2f} \rightarrow 0$ the two expressions indeed \textit{are} equivalent up to $O(z^4)$. Thus all of their divergent contributions are the same in the flat limit. However, even in this limit the two differ at order $O(z^5)$, which means that they have different finite contributions. As mentioned earlier, the finite terms we must work with are those coming from (\ref{SLam2}). An action which has both the required finite and infinite counterterms is\footnote{Note the sign of the $(\phi^3)^2$ term, which is different than the sign in (\ref{SLam2}).} 
\bea
\label{finctsa}
S_{\text{ct}} &=& \int d^5 x \sqrt{\g} \left[2 + {1 \over 4} \left(\phi^0\right)^2 + {3 \over 4} \left(\phi^3\right)^2 + 3 \sigma^2 +{1 \over 48 }\left(\phi^0\right)^4 - {3 \over 4} \(\phi^0\)^2 \sigma\right.
\no\\
&\vphantom{.}&\hspace{1.4 in}\left. + {1 \over 12} R[\g] - {1 \over 320} R[\g]^2 - {3 \over 32} R[\g] \(\phi^0\)^2 \right]
\eea
The three gravitational counterterms $2, R[\g],$ and $R[\g]^2$ match with the ones obtained in \cite{Emparan:1999pm,Alday:2014bta}. On our $S^5$ domain-wall ansatz, the term proportional to the square of the Ricci tensor simplifies in terms of the square of the Ricci scalar $R_{ij}[\gamma]R[\gamma]^{ij}=\frac15 R[\gamma]^2$. 

Note that there is still a question of curved space finite counterterms, which we have not yet fixed. If we insist on including only terms even under 
\bea
\f^0 \rightarrow -\f^0 \hspace{0.7in}\mathrm{and} \hspace{0.7in} \f^3 \rightarrow -\f^3
\eea 
(which is a symmetry of the action) it can be shown that the only way to add terms which change the curved space finite counterterms but leave the other counterterms unchanged is to add a combination of the form 
\bea
(\phi^3)^2 - {1 \over 20} R[\g] (\phi^0)^2 = 2\, e^{-f_k} \b\a \,z^5 + O(z^6)
\eea
This freedom is fixed by demanding that the vevs of the dual operators stay finite. We will simply quote the result here, \bea
\label{fincts}
S_{\text{ct}} &=& \int d^5 x \sqrt{\g} \left[2 + {1 \over 4} \left(\phi^0\right)^2 - {1 \over 2} \left(\phi^3\right)^2 + 3 \sigma^2 +{1 \over 48 }\left(\phi^0\right)^4 - {3 \over 4} \(\phi^0\)^2 \sigma\right.
\no\\
&\vphantom{.}&\hspace{1.4 in}\left. + {1 \over 12} R[\g] - {1 \over 320} R[\g]^2 - {1 \over 32} R[\g] \(\phi^0\)^2 \right]
\eea
and postpone showing that this gives finite vacuum expectation values to the next subsection.

At this level, everything has seemed unique. However, when thinking in terms of the induced fields instead of the modes appearing in asymptotic expansions, the counterterms of (\ref{fincts}) are just one of many possible sets of counterterms that can be written down. In particular, since on-shell we have the relationship 
\bea
\label{ambiguity}
I_0 \equiv 5 \sigma^2 + {45 \over 64} (\f^0)^4 - {15 \over 4} (\f^0)^2 \sigma =  O(z^6)
\eea
we may add $I_0$ freely to (\ref{fincts}) without changing either finite or infinite contributions. However, the inclusion of this term will have an impact on some of the one-point functions, which we calculate next.
%%%%%%%%%%%
\subsection{Vevs and free energy}
%%%%%%%%%%%
The renormalized on-shell action is given by 
\bea
\label{osa}
S_{\tx{ren}} = S_{\tx{6D}} + S_{\tx{GH}} + S_{\tx{ct}} + \Omega \int d^5 x~  \sqrt{\g}\, I_0
\eea
where the counterterm action $S_{ct}$ is given by (\ref{fincts}), $\Omega$ is a constant parameterizing choice of scheme, and $I_0$ is given in (\ref{ambiguity}). Note that the free energy is independent of the choice of $\Omega$, since $I_0$ is $O(z^6)$ and hence vanishes in the $\eps \rightarrow 0$ limit. However, some of the one-point functions \textit{will} depend on $\Omega$. It may be the case that only certain choices of $\Omega$ correspond to supersymmetric schemes, but since the final free energy will be independent of $\Omega$ we will not worry about this choice.

While in principle (\ref{osa}) gives us the free energy, its evaluation on our numerical solutions is complicated by the integration over $u$ in $S_{6D}$. As such, we will take a slightly roundabout approach to the calculation of the free energy, first calculating its derivative $dF/d\a$ and then integrating over the UV parameter $\a$. This will allow us to circumvent the integration over $u$. In order to get $dF/d\a$, it will first be necessary to calculate the one-point functions of the dual field theory operators. This is the topic of the following subsection.

\subsubsection{One-point functions}
By the usual AdS/CFT dictionary, the one-point functions of the operators dual to the three scalar fields and the metric are given by 
\begin{align}
\< \cO_{\sigma} \> =& \lim_{\eps \rightarrow 0}  \frac{1}{\eps^3} {1 \over \sqrt{\g}} {\delta S_{ren} \over \delta \sigma}~~~~~~~~~~
\< \cO_{\phi^0} \> = \lim_{\eps \rightarrow 0}  \frac{1}{\eps^4} {1 \over \sqrt{\g}} {\delta S_{ren} \over \delta \phi^0} 
\no\\
 \< \cO_{\phi^3} \> =& \lim_{\eps \rightarrow 0}  \frac{1}{\eps^3} {1 \over \sqrt{\g}} {\delta S_{ren} \over \delta \phi^3}~~~~~~~~~~
 \< {T^i}_j \>  = \lim_{\eps \rightarrow 0}  \frac{1}{\eps^5} {1 \over \sqrt{\g}} \gamma_{jk}{\delta S_{ren} \over \delta\gamma_{ik}}
\end{align}
We may obtain the explicit values of these vacuum expectation values by varying the on-shell action (\ref{osa}). The variation of the counterterm action $S_{ct}$ is straightforward. The variation of $S_{6D}$ gives rise to one piece which vanishes on the equations of motion, as well as a boundary term which must be accounted for. We find
\begin{align}
\label{opf1}
\< \cO_{\sigma} \> &= \lim_{\eps \rightarrow 0} {1 \over \eps^3}  \left[- 2 z \p_z \sigma + 6\sigma - {3 \over 4}( \f^0)^2 +\Omega\(10\s-\frac{15}{4} \(\phi^0\)^2\)  \right]\no \\
\< \cO_{\phi^0} \> &= \lim_{\eps \rightarrow 0} {1 \over \eps^4}  \bigg[- \frac12 \cos^2\phi^3 z \p_z \phi^0 +\frac12 \phi^0 +\frac{1}{12} \(\phi^0\)^3-\frac32 \phi^0\s-\frac{1}{16}R\phi^0 \no\\
&~~~~~~~~~~~~~~~~~~~~+\Omega\(\frac{45}{16}\(\phi^0\)^3-\frac{15}{2}\phi^0\s\)\bigg]\no \\
\< \cO_{\phi^3} \> &= \lim_{\eps \rightarrow 0} {1 \over \eps^3}  \left[ \frac12 z \p_z \phi^3-\phi^3\right]\no  \\
\< {T^i}_j\> &= \lim_{\eps \rightarrow 0} {1 \over \eps^5}\left[\frac12 \(\cK \gamma^{ij}-\cK^{ij}\)+\frac{2}{\sqrt{\gamma}}\frac{\delta S_{ct}}{\delta \gamma_{ij}}\right]
\end{align}
Evaluating the limits, we get the following one-point functions
\begin{align}
&\< \cO_{\sigma} \> = \frac52 e^{f_k}\a\b\,\Omega ~~~~~~~~~~~
\< \cO_{\phi^0} \> = \frac32 e^{-f_k} \b -\frac{15}{8} e^{f_k}\a^2\b\, \Omega
\no\\
& \< \cO_{\phi^3} \> = \frac12 \b~~~~~~~~~~~~~~~~~~~ \< {T^i}_i \>  =  -\frac52 e^{-f_k} \a\b
\label{opfAll}
\end{align}
The expectation values of the operator $\cO_{\phi^3}$ and the trace of the energy-momentum tensor are independent of $\Omega$. As a check, we note that the four one-point functions satisfy the following operator relation, which is associated to the violation of conformal invariance by non-zero classical beta functions,
\be
\langle{T^i}_i\rangle = -\sum_{\cO}(d-\Delta_{\cO})\,\phi_{\cO}\,\<\cO\> 
\ee
Here $\phi_\cO$ is the source for the operator $\cO$ and is obtained from the asymptotic solutions given in \eqref{UVexp}. 
\subsubsection{Derivative of the free energy}
Following \cite{Bobev:2013cja}, we may now compute the derivative of $F$ with respect to $\a$ as follows. First we note that
\bea
\label{dFda1}
{d F \over d \a} = {d S_{\tx{ren}} \over d\a} =\lim_{\eps \rightarrow 0} \int d^5x \sum_{\tx{fields}\,\, \Phi} {\delta \left(\sqrt{\g} \cL_{\tx{ren}}\right) \over \delta \Phi }{d \Phi \over d \a}\,\bigg|_{z = \eps}
\eea
In our case, the terms appearing in the sum over fields are
\begin{align}
\label{piece1}
{\delta \left(\sqrt{\g} \cL_{\tx{ren}}\right) \over \delta \sigma } &=  \sqrt{\g}\,  \langle O_\sigma \rangle \eps^3 + \dots~~~~~~~~~
{\delta \left(\sqrt{\g} \cL_{\tx{ren}}\right) \over \delta \phi^0 }  = \sqrt{\g} \, \langle O_\phi^0 \rangle \eps^4 + \dots
\no\\
{\delta \left(\sqrt{\g} \cL_{\tx{ren}}\right) \over \delta \phi^3 } &= \sqrt{\g}\,  \langle O_\phi^3 \rangle \eps^3+ \dots~~~~~~~~~
{\delta \left(\sqrt{\g} \cL_{\tx{ren}}\right) \over \delta \g^{ij} } = \half  \sqrt{\g}\, \langle T_{ij} \rangle  \eps^5+ \dots
\end{align}
The dots represent terms of strictly lower order in $\eps$. Furthermore, from the form of the UV asymptotic expansions (\ref{UVexp}), we have  
\begin{align}
\label{piece2}
&{d\sigma \over d \a} =  {3 \over 4}\a \eps^2 + O(\eps^3)~~~~~~~~~~~~~~~~~~~~~~~~~~~
{d\phi^0 \over d \a} = \eps + O(\eps^3)~~~~~~~~\no\\
&{d\phi^3 \over d \a} = \left(1- \a {d f_k \over d \a}\right)e^{-f_k}\eps^2 + O(\eps^3)~~~~~~~~
\frac{d\gamma^{ij}}{d\a} = -2\frac{df_k}{d\a}e^{-2f_k}\eps^2 + O(\eps^2)
\end{align}
Combining the pieces (\ref{piece1}),(\ref{piece2}) with the results for the one-point functions in (\ref{opfAll}), we find that the contribution of the metric in \eqref{dFda1} is suppressed by $\eps^2$ compared to other terms. The derivative of the free energy is then
\bea
\label{dFda}
{d F \over d \a} &=& \lim_{\eps \rightarrow 0} \int d^5 x \sqrt{\g} \, \eps^5 \left[{3 \over 2} \b e^{- f_k} + \half \b e^{- f_k}\left(1 -\a {d f_k \over d \a} \right) + O(\eps) \right]
\no\\
&=& \mathrm{vol}_0\left(S^5 \right) \, \half \b \,e^{4 f_k} \left(4 - \a {d f_k \over d \a} \right)
\eea
where $\mathrm{vol}_0(S^5)=\pi^3$ is the volume of a unit $S^5$. The $\Omega$ dependence in the one-point functions cancels out, consistent with the fact that  $F$ itself is independent of $\Omega$.
We thus obtain the final result
\bea
\label{DeFfinal}
{d F \over d \a} =  {\pi^2 \over 8\, G_6}  \b \,e^{4 f_k} \left(4-\a {d f_k \over d \a} \right)
\eea
Note that we've reintroduced the six-dimensional Newton's constant $G_6$, which had been previously set to $4 \pi G_{6} = 1$. This factor is important for the identification with the free energy on the field theory side.

\begin{figure}
\centering
\begin{minipage}{.5 \textwidth} 
\centering
\includegraphics[scale=0.8]{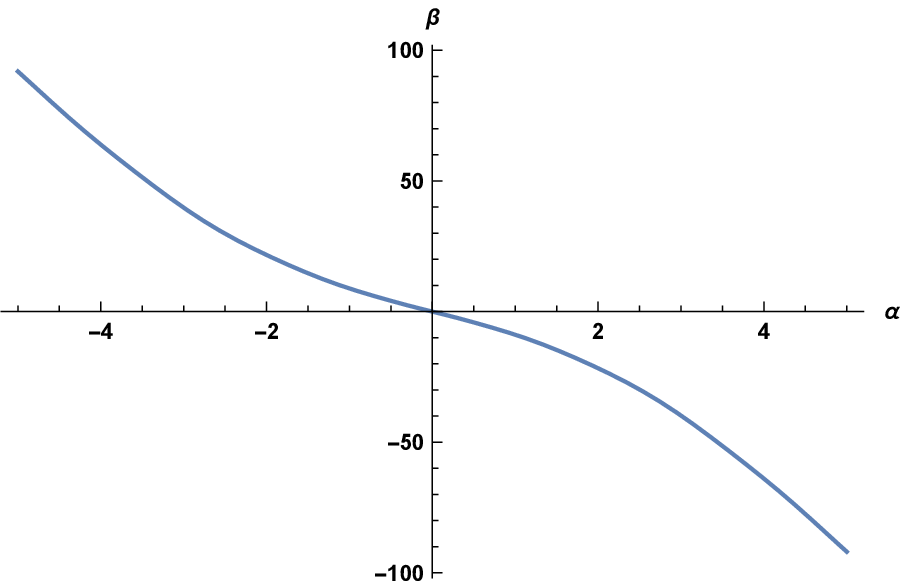}
\end{minipage}%
\begin{minipage}{.5 \textwidth} 
\centering
\includegraphics[scale=0.8]{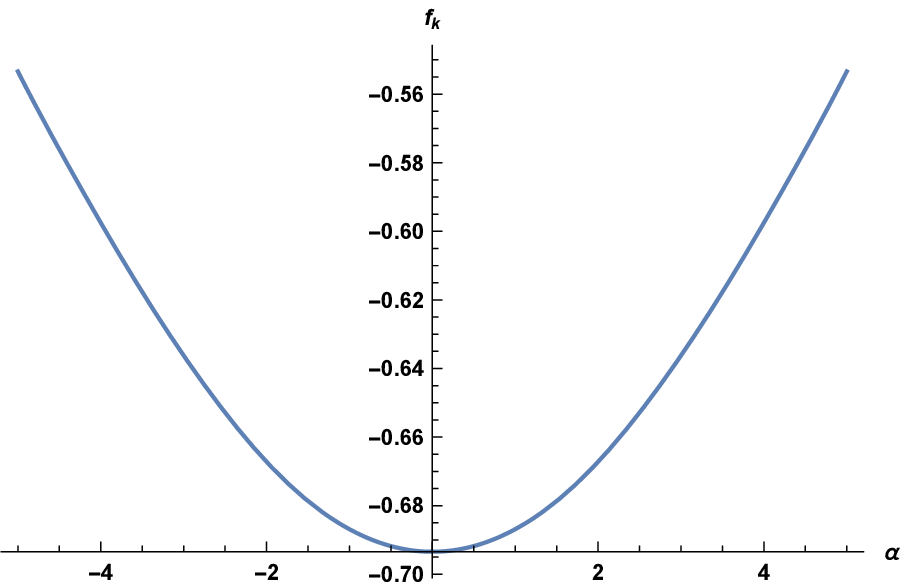}
\end{minipage}%
\caption{Plots of $\b$ vs. $\a$ and $f_k$ vs. $\a$. The relationships between the three parameters $\a,\b,$ and $f_k$ may be used to express (\ref{DeFfinal}) in terms of only a single parameter $\a$.}
\label{x33fk}
\end{figure}

\begin{figure}[h]
\centering
\includegraphics[scale=0.85]{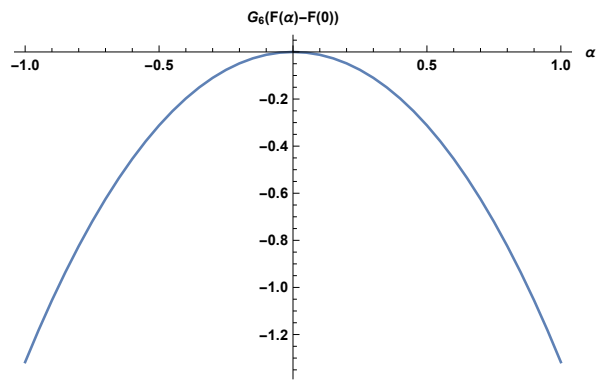}
\caption{Plot of $G_6(F(\a)-F(0))$ obtained by numerical integration of the holographic result (\ref{DeFfinal}) in the range $|\a| \leq 1$.}
\label{freeenergy1}\end{figure}

Treating $\b(\a)$ and $f_k(\a)$ as functions of $\a$, this gives us an expression which may be numerically integrated to obtain the free energy $F(\a)-F(0)$ of the domain wall. The functional forms of $\b(\a),f_k(\a)$ are obtained by fitting curves to the numerical data, as shown in Figure \ref{x33fk}. Integrating to obtain $F(\a) - F(0)$ gives the result shown in Figure \ref{freeenergy1}.

%%%%%%%%%%%%%%%%%%%%%%%%%%%%%%%%%%%%%%%%%%%%
%%%%%%%%%%%%%%%%%%%%%%%%%%%%%%%%%%%%%%%%%%%%
\section{Field theory calculation}
\setcounter{equation}{0}
\label{fieldtheory}
%%%%%%%%%%%%%%%%%%%%%%%%%%%%%%%%%%%%%%%%%%%%
%%%%%%%%%%%%%%%%%%%%%%%%%%%%%%%%%%%%%%%%%%%%

Localization \cite{Pestun:2007rz} is a powerful tool used to obtain exact results in supersymmetric quantum field theories. In the large $N$ limit, results obtained via localization calculations  can be compared with  results obtained via holography. The goal of this section is to calculate the sphere free energy for a five-dimensional mass-deformed SCFT using localization, and then to compare it to the holographic result obtained in the previous section.

A potential complication is  that the five-dimensional field theory dual to the matter-coupled six-dimensional gauged supergravity described in section \ref{sec2} has not been fully identified. This is because the full gauged supergravity has not been shown to arise as a consistent truncation of any ten-dimensional theory. In the following, the tentative field theory dual we will use for the localization calculation in the IR is a $USp(2N)$ gauge theory coupled to $N_f$ fundamental representation hypermultiplets, and a single hypermultiplet in the anti-symmetric representation. As we will review below, this theory is believed to be obtained from the D4-D8 system \cite{Brandhuber:1999np} in type I' string theory/massive type IIA supergravity.

One fundamental limitation in our comparison between field theory and holographic results is that our holographic RG flow is completely numerical, and there is no analytic formula for the free energy that can be derived from it. Nevertheless, we will find qualitative similarities between the holographic free energy and the localization result for the free energy of the aforementioned $USp(2N)$ gauge theory with mass deformation. For completeness, we will review the origin of the field theory from the brane system before presenting the localization calculation.

\subsection{The D4-D8 system}

The original D4-D8 system \cite{Brandhuber:1999np} is a brane configuration in type I' string theory involving $N$ D4 branes on $\RR^{1,8} \times S^1/\ZZ^2$. The D4 branes have their worldvolume along $\RR^{1,8}$ and sit at points along the interval $S^1/\ZZ^2$. There is an O8$^-$ plane living at each of the two ends of the interval. These orientifold planes carry $-16$ units of D8 brane charge, and thus require the inclusion of 16 D8 branes at points along the interval for tadpole cancellation. The usual construction is to stack $N_f$ D8 branes atop one of the O8$^-$ planes and to stack the remaining $(16-N_f)$ D8 branes atop the other O8$^-$ plane. One then considers the case in which the $N$ D4 branes are very near to the former stack, in which case the second boundary may be neglected. We are thus left with a consistent string theory configuration involving $N$ D4 branes probing $N_f$ D8 branes and a single O8$^-$ plane. 

This string theory setup allows for an AdS/CFT interpretation. On the closed string side of the correspondence, the near-horizon geometry of the brane configuration is found to be AdS$_6 \times S^4$ with $N$ units of 4-form flux passing through the $S^4$  \cite{Brandhuber:1999np}. This is a background of massive type IIA supergravity. While ten-dimensional uplifts of general solutions to $F(4)$ gauged supergravity are not known, pure Roman's supergravity \textit{does} have a known uplift to massive type IIA supergravity \cite{Cvetic:1999un}. In that case, the AdS$_6 \times S^4$ background may be interpreted as an AdS$_6$ background of the six-dimensional pure Roman's theory.\footnote{The reduction to six dimensions is done in two steps. One first integrates over one of the coordinates of the sphere, leaving a nine-dimensional space of the form AdS$_6 \times S^3$. Then one reduces on the $S^3$ to six dimensions, while gauging an $SU(2)$ subgroup of the sphere's $SO(4)$ isometry group \cite{Brandhuber:1999np}.} With this as motivation, we will be optimistic and assume that the solution of the six-dimensional $F(4)$ gauged supergravity theory being studied in the present case also has some uplift to massive type IIA, even though the details have not been worked out.

On the open string side of the correspondence, the worldvolume theory of the $N$ D4 branes (together with their images) is a strongly-coupled 5D SCFT which does not admit a Lagrangian description. However, one may deform this theory by a relevant operator to flow to a 5D $\cN=1$ Yang-Mills-matter theory in the IR \cite{Seiberg:1996bd}. In the setup described above, the resulting flow is to a 5D $\cN=1$ $USp(2N)$ gauge theory, where the relevant deformation has an interpretation as the gauge theory kinetic operator $ \Tr F^2$. The gauge theory is also accompanied by $N_f$ hypermultiplets in the fundamental representation and a single hypermultiplet in the antisymmetric representation. The fundamental hypermultiplets arise from D4-D8 strings, while the antisymmetric hypermultiplet arises from strings stretched between the D4 branes and their images. 

The UV SCFT has a moduli space of vacua, and this maps in the IR to the Coulomb branch of the Yang-Mills theory. The Coulomb branch is parameterized by vevs of the vector multiplet scalars, which correspond in the string theory picture to the location of the D4 branes along the interval. The locations of the D8 branes along the interval tune the masses of the fundamental hypermultiplets, while leaving the mass of the antisymmetric hypermultiplet unchanged.

From the two points of view outlined above, one is led to conjecture a duality between the fluctuations around the AdS$_6\times S^4$ background of massive type IIA supergravity on one hand, and the non-Lagrangian worldvolume theory of the $N$ D4 branes on the other. Though the non-Lagrangian nature of the field theory would naively make checking the duality extremely difficult, the fact that the UV SCFT admits a deformation to a 5D $\cN=1$ Yang-Mills theory coupled to matter allows for the following crucial simplification. Given the Lagrangian description of the IR gauge theory, we may add an infinite number of gauge-invariant, supersymmetric irrelevant operators to deform the theory back to the UV fixed point with arbitrary precision. If one assumes these irrelevant operators to be $Q$-exact, then their coefficients can be tuned freely without changing the path integral on $S^5$. Thus the sphere partition function, and hence the free energy, calculated in the IR Yang-Mills theory is expected to be equivalent to that calculated in the original non-Lagrangian theory, allowing one to test the conjectured duality. This reasoning was used in \cite{Jafferis:2012iv} to calculate the free energy on both sides of the above duality. Comparison of the two results showed a perfect match. 

We may now offer a microscopic description of the supergravity solutions described in this paper. Under the previous assumption that the solutions of the $F(4)$ gauged supergravity theory being studied here can be uplifted to an AdS$_6 \times S^4$ background of massive type IIA, our solutions should be captured by the D4-D8 brane framework. To identify the details of the relevant brane configuration, we first recall from section \ref{sec2.1} that the group which is gauged in the supergravity theory is $SU(2)_R \times G_+$, where $G_+$ is the additional gauge group arising from the presence of vector multiplets. Indeed, the presence of $n$ vector fields $A_\mu^I$ allows for the existence of a gauge group $G_+$ of dimension $\mathrm{dim} \,G_+ = n$. The gauge group $G_+$ in the bulk corresponds to a flavor symmetry group $E_{N_f+1}$ of the boundary SCFT \cite{Ferrara:1998gv}. The RG-flow triggered by the gauge coupling breaks this symmetry group to $SO(2N_f)\times U(1)$ in the IR. Deformation by the relevant mass parameters will generically break $SO(2N_f)$ further. For the solution  studied in this paper, an $SO(2)$ symmetry survives, which suggests that a minimal choice for the dual field theory would be one with $N_f=1$ (i.e. a single D8 brane).

However, even in this minimal case the enhanced gauge group $E_2 \cong SU(2) \times U(1)$ of the conformal fixed point is found to have dimension $\mathrm{dim}\, E_2=4$, which suggests that the holographic dual to such a theory should contain at least four bulk vector multiplets. Fortunately, it is possible to embed our $n=1$ solution in a theory with $n=4$, which can accommodate the extended flavor symmetry in the UV. Setting the fields of the three additional vector multiplets to vanish then reproduces exactly the solutions explored in this paper. In fact, such an embedding is possible for any value of $n>1$. This suggests that our holographic solutions are generic enough to capture the behavior of all single-mass deformations of $E_{N_f+1}$ theories for any $N_f$. As such, we will carry out the localization calculation in section \ref{massdefc} for generic $N_f$. We will find that for every choice of $1 \leq N_f \leq7$, one obtains a good match between the analytic field theory expression and our previous numerical results.

Having addressed the identification of flavor symmetries, it is natural to interpret the holographic solutions of this paper as dual to RG flows emanating from the same UV SCFTs that were found to be the duals of pure Roman's supergravity.  The flow is driven by three relevant operators of dimension $\Delta = 3,4,3$, in addition to the gauge coupling deformation which brings the non-Lagrangian UV SCFT to an IR Yang-Mills-matter theory. In the IR, the three relevant deformations are interpreted respectively as a mass term for the hypermultiplet scalars, a mass term for the hypermultiplet fermions, and a dimension three operator needed to preserve supersymmetry on the five-sphere \cite{Hosomichi:2012ek,Kallen:2012va}. The explicit form of these deformations is shown in (\ref{massdeformations}).

To support this interpretation, we now calculate the free energy of the mass-deformed $USp(2N)$ gauge theory and compare it to the holographic result displayed in Figure \ref{freeenergy1}. For the unfamiliar reader, we will first reproduce the results of \cite{Jafferis:2012iv}, where the $USp(2N)$ theory without mass deformation was studied. The techniques used for the mass-deformed theory will be the same, and the new calculation is presented in section \ref{massdefc}.

\subsection{Undeformed $USp(2N)$ gauge theory}

In \cite{Kallen:2012va}, localization techniques were used to find the perturbative partition function of $\cN = 1$ five-dimensional Yang-Mills theory with matter in a representation $R$ on $S^5$, with the result given by
\bea
\label{localizationresult}
Z &=& {1 \over |\cW|} \int_{\mathrm{Cartan}}[d\sigma] \,\,e^{- {8 \pi^3 r \over g_{YM}^2} \mathrm{Tr}(\sigma^2)}\det_{\mathrm{Ad}}\left(\sin(i \pi \sigma) e^{\half f(i \sigma)} \right)
\no\\
&\vphantom{.}&\hspace{0.6 in} \times \prod_I \det_{R_I}\left((\cos(i \pi \sigma))^{1 \over 4} e^{-{1 \over 4} f(\half - i \sigma) - {1 \over 4} f(\half + i \sigma)} \right) + O\left(e^{- 16 \pi^3 r \over g_{YM}^2}\right)
\eea
where $r$ is the radius of $S^5$, $\sigma$ is a dimensionless matrix, and $f$ is defined as 
\bea
\label{funcfdef}
f(y) = {i \pi y^3 \over 3} + y^2 \log\(1-e^{-2 \pi i y} \) + {i y \over \pi} \mathrm{Li}_2\left(e^{-2 \pi i y} \right) + {1 \over 2 \pi^2} \mathrm{Li}_3\(e^{-2 \pi i y }\) - {\zeta(3) \over 2 \pi^2}
\eea
The quotient by the Weyl group in (\ref{localizationresult}) amounts to division by a simple numerical factor $|\cW| = 2^N N!$. The integral over $\sigma$ is not restricted to a Weyl chamber. Though this localization result was obtained in the IR theory, it is expected to hold in the UV due to the assumed $Q$-exactness of the irrelevant UV completion terms. 

 One may rewrite the partition function in terms of the free energy as 
\bea
\label{partfunct2}
Z &=& {1 \over |\cW|} \int_{\mathrm{Cartan}}[d\sigma] \,e^{-F({\sigma})}+ O\left(e^{- 16 \pi^3 r \over g_{YM}^2}\right)
\no\\
F(\sigma) &=& {4 \pi^3 r \over g_{YM}^2} \mathrm{Tr}\,\sigma^2 + \mathrm{Tr}_{\mathrm{Ad}} F_V(\sigma) + \sum_I \mathrm{Tr}_{R_I} F_H(\sigma)
\eea
The definitions of $F_V(\sigma)$ and $F_H(\sigma)$ follow simply from (\ref{localizationresult}), and using (\ref{funcfdef}) one may obtain the following large argument expansions
\bea
\label{largesigexp}
F_V(\sigma) \approx {\pi \over 6} |\sigma|^3 - \pi |\sigma| \hspace{0.7 in} F_H(\sigma) \approx - {\pi \over 6} |\sigma|^3 - {\pi \over 8} | \sigma|
\eea
It was argued in \cite{Jafferis:2012iv} that in the large $N$ limit, the perturbative Yang-Mills term - i.e. the first term in the expression for $F(\sigma)$ in (\ref{partfunct2}) -  can be neglected, as can be the instanton contributions. Thus in our evaluation of the free energy, we will only concern ourselves with the contributions coming from $F_V(\sigma)$ and $F_H(\sigma)$.

The first step in the evaluation of (\ref{partfunct2}) is recasting the matrix integral in a simpler form. The integral over $\sigma$ in (\ref{partfunct2}) is an integration over the Coulomb branch, which is parameterized by the non-zero vevs of $\sigma$. One may write 
\bea
\sigma = \mathrm{diag}\{\lambda_1,\dots, \lambda_N, - \lambda_1, \dots, -\lambda_N \}
\eea
since $USp(2N)$ has $N$ elements in its Cartan. The integration variables are these $N$ $\lambda_i$. Normalizing the weights of the fundamental representation of $USp(2N)$ to be $\pm e_i$ with $e_i$ forming a basis of unit vectors for $\RR^N$, it follows that the adjoint representation has weights $\pm 2 e_i$ and $e_i \pm e_j$ for all $i \neq j$, whereas the anti-symmetric representation has only weights $e_i \pm e_j$ for all $i \neq j$. The free energy in the specific case of a vector multiplet in the adjoint, a single antisymmetric hypermultiplet, and $N_f$ fundamental hypermultiplets then is
\bea
\label{Flambdai}
F(\lambda_i) &=& \sum_{i \neq j} \left[F_V(\lambda_i - \lambda_j) + F_V(\lambda_i + \lambda_j) + F_H(\lambda_i - \lambda_j) + F_H(\lambda_i + \lambda_j) \right] 
\no\\
&\vphantom{.}&\hspace{0.3in}+ \sum_i \left[F_V(2 \lambda_i) + F_V(-2 \lambda_i) + N_f F_H(\lambda_i) + N_f F_H(-\lambda_i) \right]
\eea
The next step is to look for extrema of this function in the specific case of $\lambda_i \geq 0$ for all $i$. Extrema in the case of non-positive $\lambda_i$ can be obtained from these through action of the Weyl group.

To calculate the extrema, one first assumes that as $N \rightarrow \infty$, the vevs scale as $\lambda_i = N^\a x_i$ for $\a>0$ and $x_i$ of order $O(N^0)$. One then introduces a density function
\bea
\label{densityfunct}
\rho(x) = {1 \over N} \sum_{i=1}^N \delta(x-x_i) 
\eea
which in the continuum limit should approach an $L^1$ function normalized as 
\bea
\label{densitynorm}
\int dx\, \rho(x) = 1
\eea
In terms of this density function, one finds that 
\bea
\label{Fapprox}
F \approx -{9 \pi \over 8} N^{2 + \a} \int dx dy \,\rho(x) \rho(y) \(|x-y| + |x+y| \) + {\pi (8-N_f) \over 3} N^{1 + 3 \a} \int dx\, \rho(x)\, |x|^3
\eea
where the large argument expansions (\ref{largesigexp}) have been used, and terms subleading in $N$ have been dropped. This only has non-trivial saddle points when both terms scale the same with $N$, which demands that $\a=1/2$ and gives the famous result that $F\propto N^{5/2}$. Extremizing the free energy over normalized density functions then gives 
\bea
\label{theirfinalF}
F \approx - {9 \sqrt{2} \pi N^{5/2} \over 5 \sqrt{8-N_f}}
\eea
This value of the free energy is to be identified with the renormalized on-shell action of the supersymmetric AdS$_6$ solution. This identification yields the following relation between the six-dimensional Newton's constant $G_6$ and the parameters $N$ and $N_f$ of the dual SCFT,
\bea\label{g6rel}
G_6= \frac{5\pi\sqrt{8-N_f}}{27\sqrt{2}} ~N^{-5/2}
\eea
%%%%%%%%%%%%%%%%%%
\subsection{Mass-deformed $USp(2N)$ gauge theory}
\label{massdefc}
%%%%%%%%%%%%%%%%%
As discussed previously, we now give a mass to a single hypermultiplet in the fundamental representation. This amounts to making a shift $\sigma \rightarrow \sigma + m $ in the relevant functional determinant. The result of this shift may be accounted for in (\ref{Flambdai}) by writing
\bea
F(\lambda_i,m) &=& \sum_{i \neq j} \left[F_V(\lambda_i - \lambda_j) + F_V(\lambda_i + \lambda_j) + F_H(\lambda_i - \lambda_j) + F_H(\lambda_i + \lambda_j) \right] 
\no\\
&\vphantom{.}&\hspace{0.25in}+ \sum_i \left[F_V(2 \lambda_i)+F_V(-2 \lambda_i) +F_H(\lambda_i+m) + F_H(-\lambda_i+m) \right.
\no\\
&\vphantom{.}&\hspace{1in}\left. + (N_f-1) F_H(\lambda_i) + (N_f-1) F_H(-\lambda_i)  \right]\,\,\,\,
\eea
As before, we assume that $\lambda_i = N^\a x_i$ for $\a>0$ and introduce a density $\rho(x)$ satisfying (\ref{densitynorm}). Using the expansions (\ref{largesigexp}), we find the analog of (\ref{Fapprox}) to be
\bea
\label{ourFapprox}
F(\mu) &\approx& - {9 \pi \over 8}N^{2 + \a} \int dx dy\, \rho(x) \rho(y) \(|x-y| + |x+y| \)  + { \pi \over 3} (9-N_f)N^{1 + 3 \a} \int dx\, \rho(x) \,|x|^3
\no\\
&\vphantom{.}& \hspace{0.3 in} - {\pi \over 6} N^{1 + 3 \a} \int dx \,\rho(x) \left[ |x + \m|^3 + |x-\m|^3 \right]
\eea
where for convenience we have defined $\mu \equiv m/N^\a$. As in the undeformed case, there is a non-trivial saddle point only when $\a=1/2$.  A normalized density function which extremizes the free energy is
\bea
\rho(x) = {1 \over (8-N_f) x_*^2 - \mu^2}\left(\,2(9-N_f) |x| - |x+ \m| - |x-\m|\, \right) \hspace{0.3 in} x_* = \sqrt{9 + 2 \mu^2 \over 2(8-N_f)}\,\,\,
\eea
with $\rho(x)$ having support only on the interval $x \in [0,x_*]$. Inserting this result back into (\ref{ourFapprox}) then gives our final result,\footnote{The first term in the large $N$ expansion of this result agrees with Eq. (3.22) of \cite{Chang:2017mxc}, up to a factor of $N_f$. This difference is due to the fact that we give mass to only a single fundamental hypermultiplet. }
\bea
\label{localizationFE}
F(\mu) = {\pi \over 135}  \( (N_f-1) |\m|^5-\sqrt{2\over 8-N_f}\,(9+2 \m^2)^{5/2} \)N^{5/2}
\eea
We may check that when $\mu=0$, we reobtain the result of the undeformed case (\ref{theirfinalF}).

With this result and $G_6$ given by \eqref{g6rel}, we may now try to compare $G_6(F(\m)-F(0))$ to the same result calculated holographically in Figure \ref{freeenergy1}. Importantly, since $\m$ scales as $N^{-1/2}$, we see that in the large $N$ limit the first term of (\ref{localizationFE}) is subleading and may be neglected. Thus to leading order in $N$, the combination $G_6 F(\m)$ is in fact independent of $N_f$. Since comparison with the holographic result requires taking the large $N$ limit, our supergravity solutions will be unable to capture information about the precise flavor content of the SCFT dual. This agrees with the previous comments that, from the point of view of six-dimensional supergravity, the $n=1$ solutions we are considering can be consistently embedded into theories with any number of bulk vector multiplets.

\begin{figure}[h]
\centering
\includegraphics[scale=0.85]{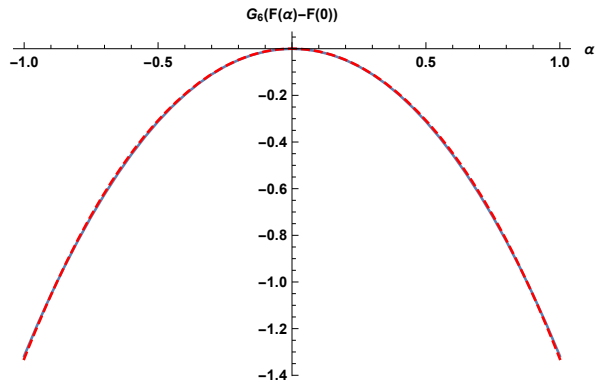}
\caption{The free energy obtained by a holographic computation (solid blue), together with the free energy obtained by a field theory localization calculation (dashed red).}
\label{freeenergy2}
\end{figure}

To proceed with the comparison between field theory and holographic results, we require a relation between the holographic deformation parameter $\a$ and the field theory mass parameter $\m$, i.e. $\a = A^{-1} \m$ for some $A$, whose numerical value can be obtained by fitting the the two results. The result of this one parameter fit is given by the red curve in Figure \ref{freeenergy2}. 

To the numerical accuracy of the holographic result, we see that the behavior of the holographic free energy as a function of the deformation parameter agrees with the field theory result obtained via localization. The value of $A$ furnishing the fit in the range $|\a| \leq 1$ is found to be $A\approx0.81$.

%%%%%%%%%%%%%%%%%%%%%%%%%%%%%%%%%%%%%%%%%%%%
%%%%%%%%%%%%%%%%%%%%%%%%%%%%%%%%%%%%%%%%%%%%
\section{Discussion }
\setcounter{equation}{0}
\label{sec5}
%%%%%%%%%%%%%%%%%%%%%%%%%%%%%%%%%%%%%%%%%%%%
%%%%%%%%%%%%%%%%%%%%%%%%%%%%%%%%%%%%%%%%%%%%

In the present paper, we used the simple setup of six-dimensional gauged supergravity coupled to a single vector multiplet to study supersymmetric mass deformations of strongly coupled five-dimensional CFTs on a five-sphere. The numerical integration of the Euclidean BPS equations and the careful treatment of holographic renormalization allowed us to obtain the holographic free energy of the theory by calculating the on-shell action for the supergravity solutions.  Due to the regularity of the solutions, the free energy depends on only one parameter, which can be interpreted as the supersymmetric mass deformation in the boundary RG flow. 

We were able to find good numerical agreement between the holographic result and a localization calculation for a free $USp(2N)$ field theory in the IR, at least in the case of reasonably small deformation parameter. This may be an example of localization working much better than expected, as we had to make unverified assumptions regarding the relation between the gauged supergravity and the underlying microscopic theory.  To understand this better, one could next consider cosets with $n>1$ and gaugings which realize larger flavor symmetries at the UV fixed points. It would also be interesting to see whether the six-dimensional solutions found here could be lifted to ten dimensions, both in the context of massive type IIA supergravity \cite{Brandhuber:1999np}
 as well as type IIB supergravity \cite{DHoker:2016ujz,DHoker:2016ysh}.
 
Furthermore, in obtaining our solutions we demanded that the five-sphere smoothly closes off in the IR.  It should also be possible to impose a different boundary condition where at finite radius one side of the RG flow is glued to a second one, resulting in a Euclidean wormhole configuration in AdS \cite{Gutperle:2002km,Maldacena:2004rf}. It is likely that such a solution would be related to the holographic defect solutions found in \cite{Gutperle:2017nwo}. We leave the study of these interesting questions for future work.

%%%%%%%%%%%%%%%%%%%%%%%%%%%%%%%%%%%%%%%%%%%%
%%%%%%%%%%%%%%%%%%%%%%%%%%%%%%%%%%%%%%%%%%%%
\section*{ Acknowledgements}

We would like to thank Christoph Uhlemann and Thomas Dumitrescu for useful conversations and Mario Trigiante and Ioannis Papadimitriou for useful email correspondence.
The work of M. Gutperle and J. Kaidi is supported in part by the National Science Foundation under grant PHY-16-19926. All authors are grateful to the Mani L.  Bhaumik Institute for Theoretical Physics for support.

\newpage
\appendix

%%%%%%%%%%%%%%%%%%%%%%%%
\section{Gamma matrix and spinor conventions }\label{gmc}
\setcounter{equation}{0}
%%%%%%%%%%%%%%%%%%%%%%%%

For concreteness, we take the following basis of gamma matrices
\bea
\gamma_1 &=\sigma_2 \otimes \mathds{1}_2 \otimes \sigma_3
\no\\
\gamma_2 &=\sigma_2 \otimes \mathds{1}_2 \otimes \sigma_1
\no\\
\gamma_3 &= \mathds{1}_2  \otimes \sigma_1  \otimes \sigma_2
\no\\
\gamma_4 &= \mathds{1}_2 \otimes \sigma_3  \otimes \sigma_2
\no\\
\gamma_5 &=\sigma_1 \otimes \sigma_2 \otimes \mathds{1}_2
\no\\
\gamma_6 &= \sigma_3 \otimes \sigma_2 \otimes \mathds{1}_2
\eea
These gamma matrices satisfy the Clifford algebra
\bea
\{ \g_\mu, \g_\nu\} = 2 \delta_{\mu \nu} 
\eea
as appropriate for a positive definite Euclidean spacetime. All matrices are purely imaginary and satisfy 
\bea
(\g_\mu)^\dagger = \g_\mu \hspace{1 in} \left(\g_\mu \right)^2 =  \mathds{1}
\eea

We will now be interested in a seven-dimensional Clifford algebra, which will require the introduction of a new matrix $\g_7$. The reason we are interested in this is that we would like to represent hyperbolic space $\mathbb{H}_6$ as a hypersurface in a seven-dimensional ambient space. This allows us to determine properties of the Dirac spinors in the Euclidean-continued $F(4)$ gauged supergravity theory with $\mathbb{H}_6$ background by first considering Dirac spinors in seven dimensions and then performing a timelike reduction. In particular, we will choose a 7D metric of signature $(+,+,+,+,+,+,-)$ for the ambient space. Then hyperbolic space $\mathbb{H}_6$ is given by the following quadratic form
\bea
\label{quadraticform}
x_1^2 + \dots +x_6^2 - x_7^2 = - L^2 
\eea 
The seven-dimensional Clifford algebra is made up of the set of matrices $\{ \g_1, \dots, \g_6, \g_7\}$, with $\g_7$ satisfying 
\bea
\label{g7props1}
(\g_7)^2 = - \mathds{1} \hspace{1 in} \{ \g_\mu, \g_7\} = 0 \,\,\,\forall \m \neq 7
\eea
As usual, we use the notation $\g^7 = (\g_7)^{-1}$, so that by the above we have $\g^7 = - \g_7$. 

We now discuss Dirac spinors in $d=7$. We define the Dirac conjugate of $\psi_A$ to be 
\bea
\bar \psi_A = \psi_A^\dagger G^{-1} 
\eea
for some matrix $G$. There are two possible choices for $G$ \cite{DAuria:2002xnk}, which in the particular case of the ambient space above are 
\bea
\label{twoGs}
G_1 = \g^7 \hspace{1 in} G_2 = \g^1 \dots \g^6
\eea
These will turn out to be the same, so we just work with the former. Thus we have that 
\bea
\label{Diracconj}
\bar \psi_A = \psi_A^\dagger \gamma_7
\eea

If we choose $\g_7$ such that 
\bea
\label{g7props2}
(\g_7)^\dagger = - \gamma_7
\eea
we can express the Hermitian conjugates of our gamma matrices as\footnote{Note that the $\eta$ used in this Appendix has nothing to do with the $\eta$ defined in (\ref{sigmaBPSeq}), though they both end up being given the value $-1$ in this paper.}
\bea
\label{etadef}
\g_\m^\dagger = \eta\, G^{-1} \g_\m G
\eea
Importantly, with $G=G_1$ in (\ref{twoGs}), we have 
\bea
\label{etares}
\eta = -1
\eea
This will be important in Appendix \ref{cosrep} when the consistency of the symplectic Majorana condition is analyzed. 
For now, we just recall that the symplectic Majorana condition must take the form 
\bea
\bar \psi_A = \epsilon^{AB} \psi_B^T\, \cC
\eea
where 
\bea
\cC^2 = 1 \hspace{0.7 in} \cC^T = \cC \hspace{0.7in} \g_\m^T = - \cC^{-1} \g_\m \cC
\eea

We now want to reduce from $d=7$ to $d=6$. In particular, we reduce on the time-like direction $x_7$. This entails finding a Euclidean induced metric on the six-dimensional surface (\ref{quadraticform}). From the point of view of the Clifford algebra, we must remove the matrix $\g_7$ to get a six-dimensional Clifford algebra. However, the properties of the matrix $\g^7$ remain the same. In fact, we may choose
\bea
\gamma_7 = \g_0\g_1\g_2\g_3\g_4\g_5
\eea
which satisfies all of the properties (\ref{g7props1}),(\ref{g7props2}).

%%%%%%%%%%%%%%%%%%%%%%%%
\section{Free differential algebra}\label{cosrep}
\setcounter{equation}{0}
%%%%%%%%%%%%%%%%%%%%%%%%
 
In this Appendix, we will construct the free differential algebra (FDA) of a supergravity theory with $\mathbb{H}_6$ background in order to motivate the form of the supersymmetry variations given in (\ref{susyvariations}).

The first step of constructing the FDA is to write down the Maurer-Cartan equations (MCEs), which may be thought of as the geometrization of the (anti-)commutation relations of the superalgebra. In short, instead of defining the algebra via the (anti-)commutators of its generators, the MCEs encode the algebraic structure in integrability conditions. In the supergravity context, a nice introduction to the MCEs, as well as to the free differential algebras to be introduced shortly, may be found in \cite{Castellani:1995gz}. In the current case, the MCEs are 
\bea
\label{MCEs}
0 &=&\cD V^a + \half \bar \psi_A \g^a \g^7 \psi^A 
\no\\
0&=& R^{ab} - 4 m^2 V^a V^b + m \bar \psi_A \g^{a b} \psi^A 
\no\\
0&=& d A^r - \half g \epsilon^{rst} A_s A_t - i \bar \psi_A \psi_B \sigma^{r\,\,AB}
\no\\
0&=& D \psi_a + m \g_a \g_7 \psi_A V^a
\eea
Here $a = 1,\dots,6$ and $V^a$ are the six-dimensional frame fields, given in terms of the seven-dimensional spin-connection as $V^a= {1 \over 2m}\omega^{a7}$. These may be compared to the analogous expressions in the dS/AdS cases of \cite{DAuria:2002xnk}.

As a simple check, the second equation of (\ref{MCEs}) tells us that when $\psi^A =0$, 
\bea
R_{\mu \nu} = - 20 m^2 g_{\mu \nu}
\eea
which is precisely as expected for an $\mathbb{H}_6$ background.

The next step is to enlarge the MCEs to a free differential algebra (FDA) by adding the following equations for the additional vector and 2-form fields of the full $d=6$ $F(4)$ supergravity theory, 
\bea
dA - m B + \a \bar \psi_A \g_7 \psi^A = 0 \hspace{1 in} d B + \b \bar \psi_A \g_a \psi^A V^a = 0
\eea
Above, $\a$ and $\b$ are two coefficients, which can be shown \cite{DAuria:2002xnk} to satisfy 
\bea
\b = - 2 \a
\eea
for our metric conventions. For the ambient space signature $(t,s) = (1,6)$, it is furthermore found that $\b = 2i$, and thus we have $\a = -i$. 

We would now like to compare the FDA above to the results of \cite{Andrianopoli:2001rs,DAuria:2000afl,DAuria:2002xnk}. To do so, we must first shift our notations by shifting 
\bea
\g^a \rightarrow \g^7 \g^a \hspace{0.7 in} \g_a \rightarrow - \g_7 \g_a
\eea
This preserves the square of the gamma matrices, and hence the signature of the metric. The definition of the Dirac conjugate spinor (\ref{Diracconj}) remains the same under this change. So the FDA for the $\mathbb{H}_6$ theory in these conventions is, 
\bea
0 &=&\cD V^a + \half \bar \psi_A \g^a \psi^A 
\no\\
0&=& R^{ab} - 4 m^2 V^a V^b + m \bar \psi_A \g^{a b} \psi^A 
\no\\
0&=& d A^r - \half g \epsilon^{rst} A_s A_t - i \bar \psi_A \psi_B \sigma^{r\,\,AB}
\no\\
0&=& D \psi_a - m \g_a \psi_A V^a
\no\\
0&=&dA - m B - i \bar \psi_A \g_7 \psi^A 
\no\\
0&=&d B -2 i \bar \psi_A \g_7 \g_a \psi^A V^a
\label{H6FDA}
\eea
We may now compare the FDA written above to that obtained in the AdS$_6$ case, which for convenience we reproduce below, 
\bea
0 &=&\cD V^a - {i \over 2} \bar \psi_A \g^a \psi^A 
\no\\
0&=& R^{ab} + 4 m^2 V^a V^b + m \bar \psi_A \g^{a b} \psi^A 
\no\\
0&=& d A^r - \half g \epsilon^{rst} A_s A_t - i \bar \psi_A \psi_B \sigma^{r\,\,AB}
\no\\
0&=& D \psi_a - i m \g_a \psi_A V^a
\no\\
0&=&dA - m B - i \bar \psi_A \g_7 \psi^A 
\no\\
0&=&d B +2  \bar \psi_A \g_7 \g_a \psi^A V^a
\eea
We see that formally, we may obtain the $\mathbb{H}_6$ FDA from the AdS$_6$ FDA by exchanging
\begin{align}
m \rightarrow - i m \qquad \psi_A \rightarrow \psi_A \qquad \bar \psi_A \rightarrow i \bar \psi_A \qquad A^r \rightarrow i A^r \qquad g \rightarrow - i g \qquad B \rightarrow - B \qquad A \rightarrow i A\no
\end{align}
These exchanges are compatible with the relation $g = 3m$. 

Finally, we will check that the $\mathbb{H}_6$ FDA is compatible with the symplectic Majorana condition. This is a statement about the fourth equation of (\ref{H6FDA}). We begin by defining 
\bea
\nabla \psi_A \equiv D \psi_A - q \g_a \psi_A V^a
\eea
where $q = m$ for $\mathbb{H}_6$ and $q=im$ for AdS$_6$. We then find that 
\bea
\overline{\nabla \psi_A}& =& D \psi_A^\dagger G^{-1} - q^* \psi_A^\dagger G^{-1} G \g_a^\dagger G^{-1} V^a = D \bar \psi_A - q^* \eta \,\bar \psi_A \g_a V^a
\no\\
\epsilon^{AB} \nabla \psi_B^T \cC &=& \epsilon^{AB} D\psi_B^T \cC - q \epsilon^{AB} \psi_B^T \cC \cC^{-1} \g_a^T \cC V^a = D \bar \psi_A+q \bar \psi_A \g_a V^a
\eea
where $\eta$ is defined implicitly in (\ref{etadef}). We thus find that the symplectic Majorana condition is consistent only when 
\bea
-q^* \eta = q
\eea
For $\mathbb{H}_6$, the consistency of the symplectic Majorana condition thus requires $\eta=-1$, which we have already seen to be the case in (\ref{etares}). On the other hand, in the AdS$_6$ case, one would instead have required $\eta=1$. Checking the results of \cite{Andrianopoli:2001rs,DAuria:2000afl} confirms that this was so.


\newpage
\begin{thebibliography}{99}

%\cite{Seiberg:1996bd}
\bibitem{Seiberg:1996bd}
  N.~Seiberg,
  ``Five-dimensional SUSY field theories, nontrivial fixed points and string dynamics,''
  Phys.\ Lett.\ B {\bf 388} (1996) 753
  %doi:10.1016/S0370-2693(96)01215-4
  [hep-th/9608111].
  %%CITATION = doi:10.1016/S0370-2693(96)01215-4;%%

%\cite{Morrison:1996xf}
\bibitem{Morrison:1996xf}
  D.~R.~Morrison and N.~Seiberg,
  ``Extremal transitions and five-dimensional supersymmetric field theories,''
  Nucl.\ Phys.\ B {\bf 483} (1997) 229
%  doi:10.1016/S0550-3213(96)00592-5
  [hep-th/9609070].
  %%CITATION = doi:10.1016/S0550-3213(96)00592-5;%%

%\cite{Intriligator:1997pq}
\bibitem{Intriligator:1997pq}
  K.~A.~Intriligator, D.~R.~Morrison and N.~Seiberg,
  ``Five-dimensional supersymmetric gauge theories and degenerations of Calabi-Yau spaces,''
  Nucl.\ Phys.\ B {\bf 497} (1997) 56
 % doi:10.1016/S0550-3213(97)00279-4
  [hep-th/9702198].
  %%CITATION = doi:10.1016/S0550-3213(97)00279-4;%%

%\cite{Brandhuber:1999np}
\bibitem{Brandhuber:1999np}
  A.~Brandhuber and Y.~Oz,
  ``The D-4 - D-8 brane system and five-dimensional fixed points,''
  Phys.\ Lett.\ B {\bf 460} (1999) 307
 % doi:10.1016/S0370-2693(99)00763-7
  [hep-th/9905148].
  %%CITATION = doi:10.1016/S0370-2693(99)00763-7;%%

%\cite{Bergman:2012kr}
\bibitem{Bergman:2012kr}
  O.~Bergman and D.~Rodriguez-Gomez,
  ``5d quivers and their AdS(6) duals,''
  JHEP {\bf 1207} (2012) 171
 % doi:10.1007/JHEP07(2012)171
  [arXiv:1206.3503 [hep-th]].
  %%CITATION = doi:10.1007/JHEP07(2012)171;%%

%\cite{Aharony:1997ju}
\bibitem{Aharony:1997ju}
  O.~Aharony and A.~Hanany,
  ``Branes, superpotentials and superconformal fixed points,''
  Nucl.\ Phys.\ B {\bf 504} (1997) 239
 % doi:10.1016/S0550-3213(97)00472-0
  [hep-th/9704170].
  %%CITATION = doi:10.1016/S0550-3213(97)00472-0;%%
  
%\cite{Aharony:1997bh}
\bibitem{Aharony:1997bh}
  O.~Aharony, A.~Hanany and B.~Kol,
  ``Webs of (p,q) five-branes, five-dimensional field theories and grid diagrams,''
  JHEP {\bf 9801} (1998) 002
 % doi:10.1088/1126-6708/1998/01/002
  [hep-th/9710116].
  %%CITATION = doi:10.1088/1126-6708/1998/01/002;%%

%\cite{DeWolfe:1999hj}
\bibitem{DeWolfe:1999hj}
  O.~DeWolfe, A.~Hanany, A.~Iqbal and E.~Katz,
  ``Five-branes, seven-branes and five-dimensional E(n) field theories,''
  JHEP {\bf 9903} (1999) 006
 % doi:10.1088/1126-6708/1999/03/006
  [hep-th/9902179].
  %%CITATION = doi:10.1088/1126-6708/1999/03/006;%%


%\cite{Nahm:1977tg}
\bibitem{Nahm:1977tg}
  W.~Nahm,
  ``Supersymmetries and their Representations,''
  Nucl.\ Phys.\ B {\bf 135} (1978) 149.
%  doi:10.1016/0550-3213(78)90218-3
  %%CITATION = doi:10.1016/0550-3213(78)90218-3;%%


%\cite{Kac:1977em}
\bibitem{Kac:1977em}
  V.~G.~Kac,
  ``Lie Superalgebras,''
  Adv.\ Math.\  {\bf 26} (1977) 8.
%  doi:10.1016/0001-8708(77)90017-2
  %%CITATION = doi:10.1016/0001-8708(77)90017-2;%%
  
  %\cite{Shnider:1988wh}
\bibitem{Shnider:1988wh}
  S.~Shnider,
  ``The Superconformal Algebra In Higher Dimensions,''
  Lett.\ Math.\ Phys.\  {\bf 16} (1988) 377.
%  doi:10.1007/BF00402046
  %%CITATION = doi:10.1007/BF00402046;%%
  
  %\cite{Passias:2012vp}
\bibitem{Passias:2012vp}
  A.~Passias,
  ``A note on supersymmetric AdS$_6$ solutions of massive type IIA supergravity,''
  JHEP {\bf 1301} (2013) 113
%  doi:10.1007/JHEP01(2013)113
  [arXiv:1209.3267 [hep-th]].
  %%CITATION = doi:10.1007/JHEP01(2013)113;%%
  
  %\cite{Lozano:2012au}
\bibitem{Lozano:2012au}
  Y.~Lozano, E.~\'{O}.~Colg{\'a}in, D.~Rodr'iguez-Go—mez and K.~Sfetsos,
  ``Supersymmetric $AdS_6$ via T Duality,''
  Phys.\ Rev.\ Lett.\  {\bf 110} (2013) no.23,  231601
%  doi:10.1103/PhysRevLett.110.231601
  [arXiv:1212.1043 [hep-th]].
  %%CITATION = doi:10.1103/PhysRevLett.110.231601;%%
  
  %\cite{Lozano:2013oma}
\bibitem{Lozano:2013oma}
  Y.~Lozano, E.~O.~O Colg‡in and D.~Rodr'guez-G—mez,
  ``Hints of 5d Fixed Point Theories from Non-Abelian T-duality,''
  JHEP {\bf 1405} (2014) 009
%  doi:10.1007/JHEP05(2014)009
  [arXiv:1311.4842 [hep-th]].
  %%CITATION = doi:10.1007/JHEP05(2014)009;%%
  
%\cite{Kelekci:2014ima}
\bibitem{Kelekci:2014ima} 
  {\"O}.~Kelekci, Y.~Lozano, N.~T.~Macpherson and E.~\'{O}.~Colg{\'a}in,
  ``Supersymmetry and non-Abelian T-duality in type II supergravity,''
  Class.\ Quant.\ Grav.\  {\bf 32}, no. 3, 035014 (2015)
  %doi:10.1088/0264-9381/32/3/035014
  [arXiv:1409.7406 [hep-th]].
  %%CITATION = doi:10.1088/0264-9381/32/3/035014;%%
  %28 citations counted in INSPIRE as of 16 Sep 2017
  
  
  %\cite{Apruzzi:2014qva}
\bibitem{Apruzzi:2014qva}
  F.~Apruzzi, M.~Fazzi, A.~Passias, D.~Rosa and A.~Tomasiello,
  %``AdS$_{6}$ solutions of type II supergravity,''
  JHEP {\bf 1411} (2014) 099
   Erratum: [JHEP {\bf 1505} (2015) 012]
  %doi:10.1007/JHEP11(2014)099, 10.1007/JHEP05(2015)012
  [arXiv:1406.0852 [hep-th]].
  %%CITATION = doi:10.1007/JHEP11(2014)099, 10.1007/JHEP05(2015)012;%%
  
  %\cite{Kim:2015hya}
\bibitem{Kim:2015hya}
  H.~Kim, N.~Kim and M.~Suh,
  ``Supersymmetric AdS$_6$ Solutions of Type IIB Supergravity,''
  Eur.\ Phys.\ J.\ C {\bf 75} (2015) no.10,  484
  %doi:10.1140/epjc/s10052-015-3705-1
  [arXiv:1506.05480 [hep-th]].
  %%CITATION = doi:10.1140/epjc/s10052-015-3705-1;%%
  
  %\cite{Kim:2016rhs}
\bibitem{Kim:2016rhs}
  H.~Kim and N.~Kim,
  ``Comments on the symmetry of AdS$_6$ solutions in string/M-theory and Killing spinor equations,''
  Phys.\ Lett.\ B {\bf 760} (2016) 780
%  doi:10.1016/j.physletb.2016.07.070
  [arXiv:1604.07987 [hep-th]].
  %%CITATION = doi:10.1016/j.physletb.2016.07.070;%%
  

  
  %\cite{DHoker:2016ujz}
\bibitem{DHoker:2016ujz}
  E.~D'Hoker, M.~Gutperle, A.~Karch and C.~F.~Uhlemann,
  ``Warped $AdS_6\times S^2$ in Type IIB supergravity I: Local solutions,''
  JHEP {\bf 1608} (2016) 046
 % doi:10.1007/JHEP08(2016)046
  [arXiv:1606.01254 [hep-th]].
  %%CITATION = doi:10.1007/JHEP08(2016)046;%%
  
  %\cite{DHoker:2016ysh}
\bibitem{DHoker:2016ysh}
  E.~D'Hoker, M.~Gutperle and C.~F.~Uhlemann,
  ``Holographic duals for five-dimensional superconformal quantum field theories,''
  Phys.\ Rev.\ Lett.\  {\bf 118} (2017) no.10,  101601
 % doi:10.1103/PhysRevLett.118.101601
  [arXiv:1611.09411 [hep-th]].
  %%CITATION = doi:10.1103/PhysRevLett.118.101601;%%
  
  %\cite{DHoker:2017mds}
\bibitem{DHoker:2017mds}
  E.~D'Hoker, M.~Gutperle and C.~F.~Uhlemann,
  ``Warped $AdS_6\times S^2$ in Type IIB supergravity II: Global solutions and five-brane webs,''
  JHEP {\bf 1705} (2017) 131
 % doi:10.1007/JHEP05(2017)131
  [arXiv:1703.08186 [hep-th]].
  %%CITATION = doi:10.1007/JHEP05(2017)131;%%
  
  
  
  
%\cite{Gutperle:2017tjo}
\bibitem{Gutperle:2017tjo}
  M.~Gutperle, C.~Marasinou, A.~Trivella and C.~F.~Uhlemann,
  ``Entanglement entropy vs. free energy in IIB supergravity duals for 5d SCFTs,''
  JHEP {\bf 1709} (2017) 125
  %doi:10.1007/JHEP09(2017)125
  [arXiv:1705.01561 [hep-th]].
  %%CITATION = doi:10.1007/JHEP09(2017)125;%%
  
  
  
  %\cite{Romans:1985tw}
\bibitem{Romans:1985tw} 
  L.~J.~Romans,
  ``The F(4) Gauged Supergravity in Six-dimensions,''
  Nucl.\ Phys.\ B {\bf 269}, 691 (1986).
  %doi:10.1016/0550-3213(86)90517-1
  %%CITATION = doi:10.1016/0550-3213(86)90517-1;%%
  %126 citations counted in INSPIRE as of 15 May 2017

  
 
  %\cite{Andrianopoli:2001rs}
\bibitem{Andrianopoli:2001rs} 
  L.~Andrianopoli, R.~D'Auria and S.~Vaula,
  ``Matter coupled F(4) gauged supergravity Lagrangian,''
  JHEP {\bf 0105}, 065 (2001)
  %doi:10.1088/1126-6708/2001/05/065
  [hep-th/0104155].
  %%CITATION = doi:10.1088/1126-6708/2001/05/065;%%
  %17 citations counted in INSPIRE as of 14 Jun 2017
  
  


  
   %\cite{Ferrara:1998gv}
\bibitem{Ferrara:1998gv} 
  S.~Ferrara, A.~Kehagias, H.~Partouche and A.~Zaffaroni,
  ``AdS(6) interpretation of 5-D superconformal field theories,''
  Phys.\ Lett.\ B {\bf 431}, 57 (1998)
  %doi:10.1016/S0370-2693(98)00560-7
  [hep-th/9804006].
  %%CITATION = doi:10.1016/S0370-2693(98)00560-7;%%
  %54 citations counted in INSPIRE as of 19 Sep 2017
  
  %\cite{DAuria:2000afl}
\bibitem{DAuria:2000afl}
  R.~D'Auria, S.~Ferrara and S.~Vaula,
  ``Matter coupled F(4) supergravity and the AdS(6) / CFT(5) correspondence,''
  JHEP {\bf 0010} (2000) 013
%  doi:10.1088/1126-6708/2000/10/013
  [hep-th/0006107].
  %%CITATION = doi:10.1088/1126-6708/2000/10/013;%%
  %36 citations counted in INSPIRE as of 13 Sep 2017
  
  
  %\cite{Karndumri:2016ruc}
\bibitem{Karndumri:2016ruc}
  P.~Karndumri and J.~Louis,
  ``Supersymmetric $AdS_6$ vacua in six-dimensional $N=(1,1)$ gauged supergravity,''
  JHEP {\bf 1701} (2017) 069
 % doi:10.1007/JHEP01(2017)069
  [arXiv:1612.00301 [hep-th]].
  %%CITATION = doi:10.1007/JHEP01(2017)069;%%
  %1 citations counted in INSPIRE as of 13 Sep 2017
  
  
  %\cite{Karndumri:2012vh}
\bibitem{Karndumri:2012vh}
  P.~Karndumri,
  ``Holographic RG flows in six dimensional F(4) gauged supergravity,''
  JHEP {\bf 1301} (2013) 134
   Erratum: [JHEP {\bf 1506} (2015) 165]
 % doi:10.1007/JHEP01(2013)134, 10.1007/JHEP06(2015)165
  [arXiv:1210.8064 [hep-th]].
  %%CITATION = doi:10.1007/JHEP01(2013)134, 10.1007/JHEP06(2015)165;%%
  
  %\cite{Karndumri:2014lba}
\bibitem{Karndumri:2014lba} 
  P.~Karndumri,
  ``Gravity duals of 5D $N=2$ SYM theory from $F(4)$ gauged supergravity,''
  Phys.\ Rev.\ D {\bf 90}, no. 8, 086009 (2014)
  %doi:10.1103/PhysRevD.90.086009
  [arXiv:1403.1150 [hep-th]].
  %%CITATION = doi:10.1103/PhysRevD.90.086009;%%
  %2 citations counted in INSPIRE as of 27 Dec 2017
  
  
  %\cite{Alday:2014rxa}
\bibitem{Alday:2014rxa}
  L.~F.~Alday, M.~Fluder, P.~Richmond and J.~Sparks,
  ``Gravity Dual of Supersymmetric Gauge Theories on a Squashed Five-Sphere,''
  Phys.\ Rev.\ Lett.\  {\bf 113} (2014) no.14,  141601
  %doi:10.1103/PhysRevLett.113.141601
  [arXiv:1404.1925 [hep-th]].
  %%CITATION = doi:10.1103/PhysRevLett.113.141601;%%
  
  
  %\cite{Alday:2014fsa}
\bibitem{Alday:2014fsa}
  L.~F.~Alday, P.~Richmond and J.~Sparks,
  ``The holographic supersymmetric Renyi entropy in five dimensions,''
  JHEP {\bf 1502} (2015) 102
  %doi:10.1007/JHEP02(2015)102
  [arXiv:1410.0899 [hep-th]].
  %%CITATION = doi:10.1007/JHEP02(2015)102;%%
  
  %\cite{Hama:2014iea}
\bibitem{Hama:2014iea}
  N.~Hama, T.~Nishioka and T.~Ugajin,
  ``Supersymmetric RŽnyi entropy in five dimensions,''
  JHEP {\bf 1412} (2014) 048
 % doi:10.1007/JHEP12(2014)048
  [arXiv:1410.2206 [hep-th]].
  %%CITATION = doi:10.1007/JHEP12(2014)048;%%
 
 %\cite{Gutperle:2017nwo}
\bibitem{Gutperle:2017nwo}
  M.~Gutperle, J.~Kaidi and H.~Raj,
  ``Janus solutions in six-dimensional gauged supergravity,''
  JHEP {\bf 1712} (2017) 018
  %doi:10.1007/JHEP12(2017)018
  [arXiv:1709.09204 [hep-th]].
  %%CITATION = doi:10.1007/JHEP12(2017)018;%%
 
 
 %\cite{Cordova:2016xhm}
\bibitem{Cordova:2016xhm}
  C.~Cordova, T.~T.~Dumitrescu and K.~Intriligator,
  ``Deformations of Superconformal Theories,''
  JHEP {\bf 1611} (2016) 135
  %doi:10.1007/JHEP11(2016)135
  [arXiv:1602.01217 [hep-th]].
  %%CITATION = doi:10.1007/JHEP11(2016)135;%%
  
 
  %\cite{Pestun:2007rz}
\bibitem{Pestun:2007rz}
  V.~Pestun,
  ``Localization of gauge theory on a four-sphere and supersymmetric Wilson loops,''
  Commun.\ Math.\ Phys.\  {\bf 313} (2012) 71
  %doi:10.1007/s00220-012-1485-0
  [arXiv:0712.2824 [hep-th]].
  %%CITATION = doi:10.1007/s00220-012-1485-0;%%

  
   %\cite{Hosomichi:2012ek}
\bibitem{Hosomichi:2012ek}
  K.~Hosomichi, R.~K.~Seong and S.~Terashima,
  ``Supersymmetric Gauge Theories on the Five-Sphere,''
  Nucl.\ Phys.\ B {\bf 865} (2012) 376
 % doi:10.1016/j.nuclphysb.2012.08.007
  [arXiv:1203.0371 [hep-th]].
  %%CITATION = doi:10.1016/j.nuclphysb.2012.08.007;%%
  
  %\cite{Kallen:2012va}
\bibitem{Kallen:2012va} 
  J.~K{\"a}ll{\'e}n, J.~Qiu and M.~Zabzine,
  ``The perturbative partition function of supersymmetric 5D Yang-Mills theory with matter on the five-sphere,''
  JHEP {\bf 1208}, 157 (2012)
 % doi:10.1007/JHEP08(2012)157
  [arXiv:1206.6008 [hep-th]].
  %%CITATION = doi:10.1007/JHEP08(2012)157;%%
  
  %\cite{Bobev:2013cja}
\bibitem{Bobev:2013cja}
  N.~Bobev, H.~Elvang, D.~Z.~Freedman and S.~S.~Pufu,
  ``Holography for $N = 2^*$ on $S^4$,''
  JHEP {\bf 1407} (2014) 001
  %doi:10.1007/JHEP07(2014)001
  [arXiv:1311.1508 [hep-th]].
  %%CITATION = doi:10.1007/JHEP07(2014)001;%%
  
  %\cite{Bobev:2016nua}
\bibitem{Bobev:2016nua}
  N.~Bobev, H.~Elvang, U.~Kol, T.~Olson and S.~S.~Pufu,
  ``Holography for $ \mathcal{N} $ = 1$^{*}$ on S$^{4}$,''
  JHEP {\bf 1610} (2016) 095
  %doi:10.1007/JHEP10(2016)095
  [arXiv:1605.00656 [hep-th]].
  %%CITATION = doi:10.1007/JHEP10(2016)095;%%
  
  
  %\cite{Freedman:2013ryh}
\bibitem{Freedman:2013ryh}
  D.~Z.~Freedman and S.~S.~Pufu,
  ``The holography of $F$-maximization,''
  JHEP {\bf 1403} (2014) 135
 % doi:10.1007/JHEP03(2014)135
  [arXiv:1302.7310 [hep-th]].
  %%CITATION = doi:10.1007/JHEP03(2014)135;%%

     %\cite{Samtleben:2008pe}
   \bibitem{Samtleben:2008pe} 
  H.~Samtleben,
  ``Lectures on Gauged Supergravity and Flux Compactifications,''
  Class.\ Quant.\ Grav.\  {\bf 25}, 214002 (2008)
  %doi:10.1088/0264-9381/25/21/214002
  [arXiv:0808.4076 [hep-th]].
  %%CITATION = doi:10.1088/0264-9381/25/21/214002;%%
  %134 citations counted in INSPIRE as of 07 Sep 2017
%  

%\cite{Lindgren:2015lia}
\bibitem{Lindgren:2015lia} 
  J.~Lindgren, I.~Papadimitriou, A.~Taliotis and J.~Vanhoof,
  ``Holographic Hall conductivities from dyonic backgrounds,''
  JHEP {\bf 1507}, 094 (2015)
  %doi:10.1007/JHEP07(2015)094
  [arXiv:1505.04131 [hep-th]].
  %%CITATION = doi:10.1007/JHEP07(2015)094;%%
  %28 citations counted in INSPIRE as of 03 Jan 2018
  
  %\cite{Cabo-Bizet:2017xdr}
\bibitem{Cabo-Bizet:2017xdr} 
  A.~Cabo-Bizet, U.~Kol, L.~A.~Pando Zayas, I.~Papadimitriou and V.~Rathee,
  ``Entropy functional and the holographic attractor mechanism,''
  arXiv:1712.01849 [hep-th].
  %%CITATION = ARXIV:1712.01849;%%

  
%  %\cite{DHoker:2008wvd}
%\bibitem{DHoker:2008wvd}
%  E.~D'Hoker, J.~Estes, M.~Gutperle, D.~Krym and P.~Sorba,
%  ``Half-BPS supergravity solutions and superalgebras,''
%  JHEP {\bf 0812} (2008) 047
%  %doi:10.1088/1126-6708/2008/12/047
%  [arXiv:0810.1484 [hep-th]].
%  %%CITATION = doi:10.1088/1126-6708/2008/12/047;%%
%  
%  
%  \bibitem{jeugt:1987}
%  J.~Can der Jeugt, ``Regular subalgebras of Lie superalgebras and extended Dynkin diagrams", J. Math. Phys {\bf 28} (1987) 292.
%  
  %\cite{Bergshoeff:2008be}
\bibitem{Bergshoeff:2008be} 
  E.~Bergshoeff, W.~Chemissany, A.~Ploegh, M.~Trigiante and T.~Van Riet,
  ``Generating Geodesic Flows and Supergravity Solutions,''
  Nucl.\ Phys.\ B {\bf 812}, 343 (2009)
  %doi:10.1016/j.nuclphysb.2008.10.023
  [arXiv:0806.2310 [hep-th]].
  %%CITATION = doi:10.1016/j.nuclphysb.2008.10.023;%%

  
  
  %\cite{Hertog:2017owm}
\bibitem{Hertog:2017owm} 
  T.~Hertog, M.~Trigiante and T.~Van Riet,
  ``Axion Wormholes in AdS Compactifications,''
  JHEP {\bf 1706}, 067 (2017)
  %doi:10.1007/JHEP06(2017)067
  [arXiv:1702.04622 [hep-th]].
  %%CITATION = doi:10.1007/JHEP06(2017)067;%% 
    
    %\cite{Ruggeri:2017grz}
\bibitem{Ruggeri:2017grz} 
  D.~Ruggeri, M.~Trigiante and T.~Van Riet,
  ``Instantons from geodesics in AdS moduli spaces,''
  arXiv:1712.06081 [hep-th].
  %%CITATION = ARXIV:1712.06081;%%
    
    
    \bibitem{Gibbons:1995vg}
  G.~W.~Gibbons, M.~B.~Green and M.~J.~Perry,
  ``Instantons and seven-branes in type IIB superstring theory,''
  Phys.\ Lett.\ B {\bf 370} (1996) 37
  %doi:10.1016/0370-2693(95)01565-5
  [hep-th/9511080].
  %%CITATION = doi:10.1016/0370-2693(95)01565-5;%%
  
  %\cite{Cortes:2003zd}
\bibitem{Cortes:2003zd}
  V.~Cortes, C.~Mayer, T.~Mohaupt and F.~Saueressig,
  ``Special geometry of Euclidean supersymmetry. 1. Vector multiplets,''
  JHEP {\bf 0403} (2004) 028
  %doi:10.1088/1126-6708/2004/03/028
  [hep-th/0312001].
  %%CITATION = doi:10.1088/1126-6708/2004/03/028;%%
  
  
%\cite{DAuria:2002xnk}
\bibitem{DAuria:2002xnk} 
  R.~D'Auria and S.~Vaula,
  ``D=6, N=2, F(4) supergravity with supersymmetric de Sitter background,''
  JHEP {\bf 0209}, 057 (2002)
  %doi:10.1088/1126-6708/2002/09/057
  [hep-th/0203074].
  %%CITATION = doi:10.1088/1126-6708/2002/09/057;%%
  %11 citations counted in INSPIRE as of 03 Dec 2017

   %\cite{Bianchi:2001kw}
\bibitem{Bianchi:2001kw} 
  M.~Bianchi, D.~Z.~Freedman and K.~Skenderis,
  ``Holographic renormalization,''
  Nucl.\ Phys.\ B {\bf 631}, 159 (2002)
  %doi:10.1016/S0550-3213(02)00179-7
  [hep-th/0112119].
  %%CITATION = doi:10.1016/S0550-3213(02)00179-7;%%
  
  %\cite{Skenderis:2002wp}
\bibitem{Skenderis:2002wp} 
  K.~Skenderis,
  ``Lecture notes on holographic renormalization,''
  Class.\ Quant.\ Grav.\  {\bf 19}, 5849 (2002)
  %doi:10.1088/0264-9381/19/22/306
  [hep-th/0209067].
  %%CITATION = doi:10.1088/0264-9381/19/22/306;%%
  
    %\cite{Papadimitriou:2004rz}
\bibitem{Papadimitriou:2004rz} 
  I.~Papadimitriou and K.~Skenderis,
  ``Correlation functions in holographic RG flows,''
  JHEP {\bf 0410}, 075 (2004)
  %doi:10.1088/1126-6708/2004/10/075
  [hep-th/0407071].
  %%CITATION = doi:10.1088/1126-6708/2004/10/075;%%
  %147 citations counted in INSPIRE as of 03 Jan 2018
  
    %\cite{Bianchi:2001de}
\bibitem{Bianchi:2001de} 
  M.~Bianchi, D.~Z.~Freedman and K.~Skenderis,
  ``How to go with an RG flow,''
  JHEP {\bf 0108}, 041 (2001)
  %doi:10.1088/1126-6708/2001/08/041
  [hep-th/0105276].
  %%CITATION = doi:10.1088/1126-6708/2001/08/041;%%
  %256 citations counted in INSPIRE as of 18 Sep 2017
  
  %\cite{Emparan:1999pm}
\bibitem{Emparan:1999pm} 
  R.~Emparan, C.~V.~Johnson and R.~C.~Myers,
  ``Surface terms as counterterms in the AdS / CFT correspondence,''
  Phys.\ Rev.\ D {\bf 60}, 104001 (1999)
  %doi:10.1103/PhysRevD.60.104001
  [hep-th/9903238].
  %%CITATION = doi:10.1103/PhysRevD.60.104001;%%
  %426 citations counted in INSPIRE as of 12 Nov 2017
 
    %\cite{Alday:2014bta}
\bibitem{Alday:2014bta} 
  L.~F.~Alday, M.~Fluder, C.~M.~Gregory, P.~Richmond and J.~Sparks,
  ``Supersymmetric gauge theories on squashed five-spheres and their gravity duals,''
  JHEP {\bf 1409}, 067 (2014)
 % doi:10.1007/JHEP09(2014)067
  [arXiv:1405.7194 [hep-th]].
  %%CITATION = doi:10.1007/JHEP09(2014)067;%%
  %20 citations counted in INSPIRE as of 12 Nov 2017
  
%\cite{Jafferis:2012iv}
\bibitem{Jafferis:2012iv}
  D.~L.~Jafferis and S.~S.~Pufu,
  ``Exact results for five-dimensional superconformal field theories with gravity duals,''
  JHEP {\bf 1405} (2014) 032
 % doi:10.1007/JHEP05(2014)032
  [arXiv:1207.4359 [hep-th]].
  %%CITATION = doi:10.1007/JHEP05(2014)032;%%
  
  %\cite{Chang:2017cdx}
\bibitem{Chang:2017cdx} 
  C.~M.~Chang, M.~Fluder, Y.~H.~Lin and Y.~Wang,
  ``Spheres, Charges, Instantons, and Bootstrap: A Five-Dimensional Odyssey,''
  arXiv:1710.08418 [hep-th].
  %%CITATION = ARXIV:1710.08418;%%
  %3 citations counted in INSPIRE as of 01 Jan 2018
  
  %\cite{Chang:2017mxc}
\bibitem{Chang:2017mxc} 
  C.~M.~Chang, M.~Fluder, Y.~H.~Lin and Y.~Wang,
  ``Romans Supergravity from Five-Dimensional Holograms,''
  arXiv:1712.10313 [hep-th].
  %%CITATION = ARXIV:1712.10313;%%
  
  %\cite{Cvetic:1999un}
\bibitem{Cvetic:1999un}
  M.~Cvetic, H.~Lu and C.~N.~Pope,
  ``Gauged six-dimensional supergravity from massive type IIA,''
  Phys.\ Rev.\ Lett.\  {\bf 83} (1999) 5226
  %doi:10.1103/PhysRevLett.83.5226
  [hep-th/9906221].
  %%CITATION = doi:10.1103/PhysRevLett.83.5226;%%
 
 %\cite{Gutperle:2002km}
\bibitem{Gutperle:2002km}
  M.~Gutperle and W.~Sabra,
  ``Instantons and wormholes in Minkowski and (A)dS spaces,''
  Nucl.\ Phys.\ B {\bf 647} (2002) 344
%  doi:10.1016/S0550-3213(02)00942-2
  [hep-th/0206153].
  %%CITATION = doi:10.1016/S0550-3213(02)00942-2;%%
  
  %\cite{Maldacena:2004rf}
\bibitem{Maldacena:2004rf}
  J.~M.~Maldacena and L.~Maoz,
  ``Wormholes in AdS,''
  JHEP {\bf 0402} (2004) 053
 % doi:10.1088/1126-6708/2004/02/053
  [hep-th/0401024].
  %%CITATION = doi:10.1088/1126-6708/2004/02/053;%%
  
  
  %\cite{Castellani:1995gz}
\bibitem{Castellani:1995gz} 
  L.~Castellani and A.~Perotto,
  ``Free differential algebras: Their use in field theory and dual formulation,''
  Lett.\ Math.\ Phys.\  {\bf 38}, 321 (1996)
  %doi:10.1007/BF00398356
  [hep-th/9509031].
  %%CITATION = doi:10.1007/BF00398356;%%
  %8 citations counted in INSPIRE as of 05 Dec 2017
 
  

  
\end{thebibliography}
\end{document}